\documentclass[a4paper,11pt]{article}
\usepackage{jcappub}
\usepackage[utf8]{inputenc}
\usepackage{amsmath,mathtools}

\usepackage{tensor}
\usepackage{amssymb}
\usepackage{graphicx}
\usepackage{physics}
\usepackage{lipsum}  
\usepackage{bigints}
\usepackage{verbatim}
\usepackage[dvipsnames]{xcolor}
\newcommand{\leri}[1]{\left(#1\right)}

\newcommand{\frstheta}{\tensor[^{(1)}]{\theta}{}(x)}

\usepackage{subcaption}

\title{Quasinormal modes of Schwarzschild black holes in projective invariant Chern-Simons modified gravity}

\author[a,b,1]{S. Boudet,\note{Corresponding author.}}
\author[c]{F. Bombacigno,}
\author[c,d]{Gonzalo J. Olmo}
\author[e]{P. J. Porf\'{i}rio}

\affiliation[a]{Dipartimento di Fisica, Universit\`{a} di Trento,\\Via Sommarive 14, I-38123 Povo (TN), Italy}
\affiliation[b]{Trento Institute for Fundamental Physics and Applications (TIFPA)-INFN,\\Via Sommarive 14, I-38123 Povo (TN), Italy}
\affiliation[c]{Departament de F\'{i}sica Teòrica and IFIC, Centro Mixto Universitat de València - CSIC, Universitat de València, Burjassot 46100, València, Spain}
\affiliation[d]{Universidade Federal do Cear\'a (UFC), Departamento de F\'isica,\\ Campus do Pici, Fortaleza - CE, C.P. 6030, 60455-760 - Brazil.}
\affiliation[e]{Departamento de F\'{\i}sica, Universidade Federal da 
	Para\'{\i}ba,\\
	Caixa Postal 5008, 58051-970, Jo\~ao Pessoa, Para\'{\i}ba, Brazil}

\emailAdd{simon.boudet@unitn.it}
\emailAdd{flavio2.bombacigno@uv.es}
\emailAdd{gonzalo.olmo@uv.es}
\emailAdd{pporfirio@fisica.ufpb.br}

\abstract{
We generalize the Chern-Simons {modified gravity} to the metric-affine case and impose projective invariance by supplementing the Pontryagin density with homothetic curvature terms which do not spoil topologicity. The latter is then broken by promoting the coupling of the Chern-Simons term to a {(pseudo)}-scalar field. The solutions for torsion and nonmetricity are derived perturbatively, showing that they can be iteratively obtained from the background fields. This allows us to describe the dynamics for the metric and the scalar field perturbations in a self-consistent way, and we apply the formalism to the study of quasinormal modes in a Schwarzschild black hole background. Unlike in the metric formulation of this theory, we show that the scalar field is endowed with dynamics even in the absence of its kinetic term in the action. Finally, using numerical methods we compute the quasinormal frequencies and characterize the late-time power law tails for scalar and metric perturbations, comparing the results with the outcomes of the purely metric approach.}

\makeatletter
\gdef\@fpheader{}
\makeatother

\begin{document}



\maketitle
\flushbottom

\section{Introduction}
\noindent The study of extended theories of gravity is aimed to deepen our understanding of physical phenomena that still miss a coherent explanation within General Relativity (GR). In the last years, several alternatives have been proposed, each tackling one or more open problems in cosmology and astrophysics \cite{Sotiriou:2008rp,Nojiri:2010wj,Cai:2015emx,NOJIRI20171,Krssak:2018ywd,Olmo:2019flu,Cabral:2020fax,Harko:2020ibn,Capozziello:2022lic,Fernandes:2022zrq}. When dealing with black hole physics, however, to select a peculiar class of modified theories can result in a non-trivial task. It is known, in fact, that Schwarzschild and Kerr solutions are actually common, at least at the kinematic level, to a wide number of alternative formulations, and only a dynamical characterization can provide strong field tests for observational constraints. It can be more instructive, therefore, to describe deviations from GR by a suitable set of parameters, quantifying the departure of the metric from standard solutions, and requiring that Schwarzschild and Kerr scenarios could be reproduced continuously in some adequate limit. 

This idea is pursued, for instance, in \cite{PhysRevD.83.124015}, where modifications to the Kerr solution are implemented in a rather natural way for arbitrary values of spin, by means of a stationary, axisymmetric and asymptotically flat metric describing generic rotating solutions in modified gravity frameworks, irrespective of the actual theory generating them. This approach has the benefit of overcoming most of the pathologies usually arising in perturbation theory within GR context, mainly due to the violation of no-hair theorems \cite{Gair:2007kr,Johannsen:2010xs,Sotiriou:2015pka,Herdeiro:2014goa,Herdeiro:2015waa,Cardoso:2016ryw}, which severely restrict predictions to specific cases, as for instance extreme mass-ratio inspiraling \cite{Barack:2006pq} or motion of stars and pulsars around black holes \cite{Wex:1998wt,Will:2007pp,Broderick:2013rlq}. The absence, however, of a concrete model responsible for the appearance of the metric proposed in \cite{PhysRevD.83.124015}, prevents its extension to different settings other than black holes, raising the issue about how this kind of solution could be actually obtained.

In this respect, Chern-Simons  {modified gravity}  {(CSMG)} stands as an interesting candidate, able to address quantum gravity and black hole problems within a unique theoretical setting.  {The CSMG was originally proposed by Jackiw and Pi \cite{Jackiw:2003pm} inspired by the Chern-Simons modification of  electrodynamics \cite{Carroll:1989vb}. As for the U(1) gauge theory, the Maxwell Lagrangian is modified by introducing a (pseudo)-scalar field, $\theta(x)$, coupled to the U(1)-gauge topological Pontryagin density, $\,^{*}FF\equiv \,^{*}F^{\mu\nu}F_{\mu\nu}$. Even maintaining the gauge invariance, such a modification allows for Lorentz/CPT symmetry violation \cite{Carroll:1989vb}.
It can be readily seen by casting the modified term into the Carroll-Field-Jackiw form, i.e., $v_{\mu}\,^{*}F^{\mu\nu}A_{\nu}$, where $v_{\mu}\equiv \partial_{\mu}\theta$ is an axial vector responsible for Lorentz/CPT symmetry breaking. In a wider picture, $v_{\mu}$ corresponds to one of the coefficients for Lorentz/CPT violation in the Standard Model Extension (SME) \cite{Colladay:1996iz, Colladay:1998fq, Kostelecky:2003fs}. In much the same way as in the 
Chern-Simons modification of Maxwell electrodynamics, CSMG adds to the  Lagrangian of GR a non-minimal coupling between $\theta(x)$ and the gravitational Pontryagin density which, in turn, is defined by $\,^{*}RR\equiv\,^{*}R^{\mu\nu\alpha\beta}R_{\nu\mu\alpha\beta}$.}

The theory has been explored in two different scenarios, known as non-dynamical and dynamical cases. In the former, the pseudo-scalar
field is interpreted as a purely external quantity. In the latter, it possesses dynamics, satisfying
its own equation of motion. Both approaches provide modified field equations as compared to those of GR. In the non-dynamical framework, 
the contributions stemming from the CS curvature term are encoded by the C-tensor, sometimes called Cotton tensor \cite{Jackiw:2003pm, Alexander:2009tp}. Furthermore, the
modified field equations fulfill the Pontryagin constraint, $\,^{*}RR=0$, in order to guarantee the diffeomorphism invariance of the theory.
On the other hand, the situation changes drastically in the dynamical framework. Firstly, because as a dynamical variable, $\theta(x)$ provides a new contribution to the metric equation via its stress-energy tensor, which comes in addition to the C-tensor. Secondly, because the scalar field satisfies
a dynamical equation of motion instead of the Pontryagin constraint.

The relevance of CSMG in addressing the issue raised by \cite{Johannsen:2010xs}, can be then appreciated by looking at the symmetry properties of the Pontryagin density under parity transformations. The requirement that the CS correction be parity preserving, implies that the scalar coupling must be mediated by a pseudo-scalar field (parity odd), in agreement with its hypothetical string theory origin. This, in turn, implies that parity-even solutions, such as the Schwarzschild black hole  {and, more generically, spherically symmetric} metrics remain unaffected by Chern-Simons corrections, which only manifest themselves in parity violating scenarios, i.e. in Kerr-type or rotating spacetimes \cite{PhysRevD.79.084043}. In other words, the CS term can spontaneously generate Kerr deformations and offer a theoretical justification to \cite{Johannsen:2010xs}.  {Other non-trivial solutions have been obtained in literature also for rotating G\"{o}del-type space-times \cite{our1, our2, our3, our4} or Einstein-dilaton Gauss-Bonnet gravity \cite{Kanti:1995vq,Kanti:1997br,Kleihaus:2011tg,Ayzenberg:2014aka,Maselli:2015tta,Kleihaus:2015aje,Okounkova:2019zep,Cano:2019ore,Delgado:2020rev,Pierini:2021jxd}. The remarkable impact of the CS term in those solutions turned out to be
primordial to find a large class of completely causal solutions in a distinguishing way to GR.}

The role of CSMG in the arena of modified gravity theories is also strongly motivated by several arguments arising from different physical backgrounds, where the presence of a CS term seems to be ubiquitous (see \cite{Alexander:2009tp} for a review). In particle physics, for instance, the gravitational anomaly turns out to be proportional to the Pontryagin density, and a CS-like counterterm must be included in the action to cancel the anomaly out. Counterterms of this kind can be actually produced also in string theory via the Green-Schwarz mechanism and emerge in low energy effective string models \cite{PhysRevD.77.024015,Adak2012}. Remarkably, some analogies can be outlined with loop quantum gravity approaches \cite{Ashtekar:1988sw} as well, where CS corrections arise in addressing chiral anomaly of fermions and the Immirzi field ambiguity \cite{Perez:2005pm,Freidel:2005sn,Date:2008rb,Mercuri:2009zi,Mercuri:2009vk}. Moreover, this theory may help in designing new strategies to {probe the (local) Lorentz/CPT} symmetry breaking in gravitation, which is expected to receive new observational inputs in the next few years. Indeed, CS parity violation effects are already well established in contexts such as amplitude birefringence for gravitational wave propagation \cite{Jackiw:2003pm,Martin-Ruiz:2017cjt,Nojiri:2019nar,Nojiri:2020pqr,NOJIRI2020100514}, CMB polarization \cite{ALEXANDER2008444,PhysRevLett.83.1506,Bartolo:2018elp,Bartolo:2017szm} and the baryon asymmetry problem \cite{PhysRevLett.96.081301,PhysRevD.69.023504,Alexander_2006}.

 As is well known, geometric theories of gravity can always be formulated following different approaches, depending on the a priori assumptions about their metric and affine structures. When the metric and the connection are a priori independent, one refers to the metric-affine formulation, while when the connection is forced to be compatible with the metric, one speaks of the usual metric approach. The two formulations are generally not equivalent and in most cases they actually lead to two completely different theories, as is also the case for CSMG. In this respect, the metric-affine version is closer to a gauge theory than the purely metric one \cite{Hehl:1994ue,Zanelli:2005sa} and since the CS term is constructed using the connection of the corresponding gauge field, it is rather surprising that the literature on CSMG has mainly focused on the study of its metric version, while the metric-affine formulation has received only timid attention \cite{Deser:2006ht}. From another perspective, the first-order CSMG has been discussed within the Cartan formalism  \cite{doi:10.1063/1.529191,PhysRevD.64.084012,CACCIATORI20062523,PhysRevD.78.025002,Cisterna2019}, with a focus on theoretical aspects, while the derivation of observable effects has received little attention. 
 
 In this work we will be dealing with the metric-affine formulation of {CSMG}, and we will neglect the coupling of matter with connection, i.e. we will assume a vanishing hypermomentum (see \cite{Kranas:2018jdc,Iosifidis:2020gth,Iosifidis:2021fnq}). Instead of retaining the usual expression for the CS term and simply promoting the connection to an independent variable, we will also require the invariance of the theory under projective transformations, which results in the inclusion of an additional term in the action (see \cite{PhysRevD.103.124031} for analogous studies concerning the Nieh-Yan topological invariant). Projective transformations consist on a shift in the independent connection \cite{Afonso:2017bxr,Iosifidis:2019fsh}, and the breaking of its associated symmetry has been shown to be related to the emergence of dynamical instabilities in some classes of metric-affine theories \cite{BeltranJimenez2019}. It is therefore sensible to take care of the projective symmetry when promoting the theory to the metric-affine formalism. 
 
A natural difficulty that one faces when dealing with metric-affine theories is the need to solve for the connection equations. Though formal exact solutions can be found in some cases, like in Ricci-based gravity theories \cite{Afonso:2017bxr} and models with non-minimal couplings \cite{Delhom:2022xfo,Delhom:2020gfv,Delhom:2019wcm}, the structure of the corresponding equations in CSMG makes this task rather challenging, forcing us to consider a perturbative approach. For this reason, and because the study of black hole perturbations \cite{Maggiore:2018sht} is of special interest given current and future observational capabilities, here we focus on  
 the simple scenario of perturbations around the Schwarzschild background solution. 
 
 A standard decomposition in tensor harmonics allows for a clear separation of the metric perturbation in even and odd parity modes. Their evolution is then described by two differential equations, the Zerilli and the Regge-Wheeler equations, where the features of the background spacetime are encoded in an effective potential term. Several methods for their solution exist in  the literature, showing how the evolution of metric perturbations is mainly characterized by quasinormal modes (QNMs), consisting of damped oscillations, and a late-time power law tail. The initial oscillatory behavior is characterized by complex frequencies with a negative imaginary part and it represents the black hole response to an initial perturbation, due for instance to infalling matter or characterizing the nonequilibrium configuration of a black hole immediately after its formation by a merging event. On the other hand, the late-time tail only depends on the spacetime properties at spacial infinity, namely on the asymptotic form of the effective potential.
 
Besides metric tensor perturbations, one can also consider perturbations of a scalar field on the Schwarzschild background. In GR, i.e. for a Klein-Gordon scalar field on curved background, they are characterized by the same late-time behaviour as tensor perturbations, but with a different set of quasinormal frequencies, and are decoupled from tensor ones. This is no longer the case when considering perturbations of scalar fields nonminimally coupled to the geometry, as in the case of {CSMG}, where the scalar field couples to the Pontryagin density. The study of black hole QNMs in metric {CSMG} has been carried out in several works \cite{PhysRevD.80.064008,PhysRevD.81.089903,PhysRevD.81.124021,PhysRevD.80.064006}, though the metric-affine version has not been explored yet. In Sec.~\ref{sec: metric CS comparison} we shall discuss the conclusions obtained in the metric formulation in comparison with our results in the metric-affine case. In this sense, it is worth noting that in the context of black hole QNMs the non-dynamical metric {CSMG}  has been shown to have several shortcomings, such as being an overconstrained system of equations, while the dynamical theory seems viable. As we will see, in the metric-affine case both choices actually lead to a dynamical theory, even in the absence of a kinetic term for the scalar field in the action.

The paper is organized as follows. In section \ref{sec2} we introduce the projective invariant generalization of {CSMG} to the metric-affine case, discussing in some detail how it is related to topologicity. In section \ref{sec3} we derive the equation for the affine connection, showing how it can be simplified by taking advantage of projective symmetry. In section \ref{sec4} we assume for the scalar field a perturbative expansion, and we evaluate at the linearized level the explicit solution for the connection, then in section \ref{sec5} we calculate on half shell the equations for the metric and scalar perturbations. In \ref{sec6} we deal with the QNMs of Schwarzschild black holes, and we discuss the properties of the model according the value of the parameter $\beta$ ruling the scalar field kinetic term in the action, comparing our results with the purely metric case. Finally, in section \ref{sec7} conclusions are drawn and future perspectives outlined.

Spacetime signature is chosen mostly plus and the gravitational coupling set as $\kappa^2=8\pi$, using geometrized units: $G=c=1$. The Levi-Civita tensor $\varepsilon_{\mu\nu\rho\sigma}$  is defined in terms of the completely antisymmetric symbol $\epsilon_{\mu\nu\rho\sigma}$, with $\epsilon_{0123}=1$. For details on the metric-affine formalism and the conventions used we refer the reader to App.~\ref{appendix A}.

\section{Projective invariant generalization of the Chern-Simons term}\label{sec2}
\noindent In the standard metric formulation, the dynamical {CSMG} is described by \cite{Alexander:2009tp}
\begin{equation}
S=\frac{1}{2\kappa^2}\int d^4x\sqrt{-g}\leri{R+\frac{\alpha}{8}\theta(x)\varepsilon\indices{^{\mu\nu\rho\sigma}}R\indices{^\alpha_{\beta\mu\nu}}R\indices{^\beta_{\alpha\rho\sigma}} -\frac{\beta}{2}\nabla_\mu \theta \nabla^\mu\theta-\beta V(\theta)},
\label{action CS met}
\end{equation}
where $\alpha$ and $\beta$ are two real parameters, ruling the Pontryagin density correction to the Einstein-Hilbert action and the kinetic term for the scalar field $\theta(x)$, respectively. For the sake of completeness we also include a potential term $V(\theta)$. When $\theta(x)$ boils down to a constant, the Pontryagin term simply reduces to a boundary contribution (see the seminal paper \cite{Jackiw:2003pm} or \cite{Alexander:2009tp}) and we recover General Relativity. Recall, as pointed out in the introduction, that the scalar field equation of the non-dynamical framework, $\beta=0$, boils down to a constraint for the allowed metrics, while in the dynamical case, $\beta\neq 0$, they represent a new independent degree of freedom. As such, there is no continuous transition in the limit $\beta\to 0$. This can also be traced back to the fact that the corrections to the metric tensor in the dynamical case are proportional to $\alpha^{2}/\beta$, and are ill defined as $\beta\to 0$ \cite{Yunes:2009hc}.

We are interested, however, in a metric-affine generalization of \eqref{action CS met}, where the Riemann tensor is defined starting from the independent connection, i.e:
\begin{equation}
    \mathcal{R}\indices{^\rho_{\mu\sigma\nu}}=\partial_\sigma\Gamma\indices{^\rho_{\mu\nu}}-\partial_\nu\Gamma\indices{^\rho_{\mu\sigma}}+\Gamma\indices{^\rho_{\tau\sigma}}\Gamma\indices{^\tau_{\mu\nu}}-\Gamma\indices{^\rho_{\tau\nu}}\Gamma\indices{^\tau_{\mu\sigma}},
\end{equation}
which we are considering as general as possible, so that in principle torsion and nonmetricity tensors are non vanishing and defined by:
\begin{equation}
    \begin{split}
        &T\indices{^\rho_{\mu\nu}}\equiv\Gamma\indices{^\rho_{\mu\nu}}-\Gamma\indices{^\rho_{\nu\mu}}, \\
        &Q\indices{_{\rho\mu\nu}}\equiv-\nabla_\rho g_{\mu\nu}.
    \end{split}
\end{equation}
where $\nabla_\mu$ is the covariant derivative built from the entire connection. Then, in metric-affine scenarios we can introduce the notion of projective transformations, given by the connection shift
\begin{equation}
    \Gamma\indices{^\rho_{\mu\nu}}\rightarrow\tilde{\Gamma}\indices{^\rho_{\mu\nu}}=\Gamma\indices{^\rho_{\mu\nu}}+\delta\indices{^\rho_\mu}\xi_\nu,
    \label{projective}
\end{equation}
where $\xi_\mu$ is an a priori undefined one-form. It is easy to show that the Ricci scalar is invariant\footnote{The reader is referred to Appendix~\ref{appendix A} for all details concerning the metric-affine formulation of theories of gravity.} under \eqref{projective}, so that projective transformations can be considered in the Palatini formulation of GR as an additional gauge symmetry of such an enlarged kinematic structure (see \cite{Hehl:1994ue} for details). However, when higher order corrections are included in the action, projective invariance is not assured and its explicit violation could lead to dynamical instabilities \cite{BeltranJimenez2019}. Now, since for \eqref{projective} the Riemann tensor transforms as
\begin{equation}
    \mathcal{R}\indices{^\rho_{\mu\sigma\nu}}\rightarrow\tilde{\mathcal{R}}\indices{^\rho_{\mu\sigma\nu}}=\mathcal{R}\indices{^\rho_{\mu\sigma\nu}}-\delta\indices{^\rho_\mu}\partial_\sigma\xi_\nu+\delta\indices{^\rho_\mu}\partial_\nu\xi_\sigma,
    \label{projective transformation Riemann}
\end{equation}
it is easy to show that the Pontryagin density is not invariant for generic projective transformations, unless we can express the one-form $\xi_\mu$ as the gradient of a scalar field $\Phi$, i.e. $\xi_{\mu}=\partial_{\mu}\Phi$, which are called special projective transformations. This implies that a projective invariant formulation of {CSMG} cannot be simply obtained by rewriting \eqref{action CS met} in metric-affine formalism, but some additional terms in curvature are in general expected. We consider, therefore, the following generalization
\begin{equation}
    \mathcal{CS}\equiv\epsilon^{\mu\nu\rho\sigma}\leri{\mathcal{R}\indices{^\alpha_{\beta\mu\nu}}+\delta\indices{^\alpha_\beta}\leri{a_1\, \mathcal{R}_{\mu\nu}+b_1\,\hat{\mathcal{R}}_{\mu\nu}}}\leri{\mathcal{R}\indices{^\beta_{\alpha\rho\sigma}}+\delta\indices{^\beta_\alpha}\leri{a_2\, \mathcal{R}_{\rho\sigma}+b_2\,\hat{\mathcal{R}}_{\rho\sigma}}},
    \label{CS general factorized}
\end{equation}
where, without loss of generality, we set to unity a possible coefficient in front of the Riemann tensor, which can be always reabsorbed in the scalar field definition, and we also introduced a homothetic and Ricci tensor curvature contribution (see Appendix~\ref{appendix A}).
Now, taking into account the transformation properties of the tensor quantities involved in \eqref{CS general factorized}, it can be demonstrated that projective invariance is recovered only for 
\begin{equation}
    1+a_i+4b_i=0,\;i=1,2.
    \label{condition proj inv}
\end{equation}
By solving for $b_i=-\frac{a_i+1}{4}$, we can rearrange the projective invariant CS term as
\begin{equation}
    \mathcal{CS}\equiv\epsilon^{\mu\nu\rho\sigma}\leri{\mathcal{R}\indices{^\alpha_{\beta\mu\nu}}+\delta\indices{^\alpha_\beta}\leri{a_1\, \mathcal{R}_{\mu\nu}-\frac{a_1+1}{4}\hat{\mathcal{R}}_{\mu\nu}}}\leri{\mathcal{R}\indices{^\beta_{\alpha\rho\sigma}}+\delta\indices{^\beta_\alpha}\leri{a_2\, \mathcal{R}_{\rho\sigma}-\frac{a_2+1}{4}\hat{\mathcal{R}}_{\rho\sigma}}}.
    \label{CS general factorized 2}
\end{equation}
We observe that we could have also started, in principle, from a CS form of the type
\begin{equation}
    \mathcal{CS}^*\equiv\epsilon^{\mu\nu\rho\sigma}\leri{\mathcal{R}\indices{^\alpha_{\beta\mu\nu}}\mathcal{R}\indices{^\beta_{\alpha\rho\sigma}}+A\, \mathcal{R}_{\mu\nu}\mathcal{R}_{\rho\sigma}+B\,\hat{\mathcal{R}}_{\mu\nu}\hat{\mathcal{R}}_{\rho\sigma}+C\,\mathcal{R}_{\mu\nu}\hat{\mathcal{R}}_{\rho\sigma}},
    \label{CS general unfactorized}
\end{equation}
and in this case demanding projective invariance would have led to the following set of constraints:
\begin{equation}
    B=\frac{A-4}{16},\;C=-\frac{A}{2},
    \label{condition proj inv unfact}
\end{equation}
where, with respect to \eqref{CS general factorized}, we have only one free parameter. Such a discrepancy is due to the fact that \eqref{CS general factorized} can be always rearranged into the form \eqref{CS general unfactorized}, up to the redefinition
\begin{equation}
    A=4a_1 a_2,\;B=\frac{a_1a_2-1}{4},\;C=-2a_1a_2,
\end{equation}
which is easy to see to satisfy \eqref{condition proj inv unfact}, whereas the converse does not hold in general, since \eqref{CS general unfactorized} cannot be always factorized as in \eqref{CS general factorized}.

Now, if we also require that such a modified CS term could be still expressed as the divergence of a quadricurrent, i.e. a boundary term, we are compelled to further restrict the parameter space to $A=C=0$, which simply leaves us with $B=-\frac{1}{4}$ (equivalently $a_1=0$ or $a_2=0$). When we include  contractions with the Ricci tensor as in \eqref{CS general unfactorized}, calculations show that topologicity is explicitly broken, in contrast with the homothetic part, which can be still rearranged as 
\begin{equation}
   \epsilon^{\mu\nu\rho\sigma}\hat{\mathcal{R}}_{\mu\nu}\hat{\mathcal{R}}_{\rho\sigma}=\epsilon^{\mu\nu\rho\sigma}\partial_\mu\leri{\Gamma\indices{^\alpha_{\alpha\nu}}\partial_\rho\Gamma\indices{^\beta_{\beta\sigma}}} \ .
\end{equation}
Similarly to what occurs in the metric-affine generalization of the Nieh-Yan term (see \cite{PhysRevD.103.124031}), also for the Chern-Simons case projective invariance is not directly related to topologicity, and, in principle, we can always violate the former and still retain the latter by choosing $B\neq -\frac{1}{4}$. 

\section{Connection in projective invariant CSMG}\label{sec3}
\noindent In our analysis we simply set $B=-\frac{1}{4}$, so that we assume as the starting point of our work the projective invariant generalization of \eqref{action CS met}, i.e.
\begin{equation}
S=\frac{1}{2\kappa^2}\int d^4x\sqrt{-g}\leri{\mathcal{R}+\frac{\alpha}{8}\theta(x)\varepsilon\indices{^{\mu\nu\rho\sigma}}\leri{\mathcal{R}\indices{^\alpha_{\beta\mu\nu}}\mathcal{R}\indices{^\beta_{\alpha\rho\sigma}}-\frac{1}{4}\hat{\mathcal{R}}\indices{_{\mu\nu}}\hat{\mathcal{R}}\indices{_{\rho\sigma}}} -\frac{\beta}{2}\nabla_\mu \theta \nabla^\mu\theta},
\label{action CS}
\end{equation}
where we omit the potential term for the scalar field $\theta(x)$. Like in the metric formulation, though this action depends explicitly on two parameters $\alpha$ and $\beta$, one of them can be absorbed into a redefinition of $\theta$. This will be useful in Sec.~\ref{sec6}, where we will set $\alpha\to 1$.

Now, since the modified Chern-Simons term can be still rewritten as a total derivative containing cubic contributions in the affine connection, we expect in general that the resulting equation of motion for $\Gamma\indices{^\lambda_{\mu\nu}}$ be affected by second order terms, which can introduce substantial differences as compared to the standard Levi-Civita solution. In particular, by varying \eqref{action CS} with respect to the affine connection, we obtain
\begin{align}
    &-\nabla_\lambda\leri{\sqrt{-g} g^{\mu\nu}}+\delta^\nu_\lambda\nabla_\rho\leri{\sqrt{-g} g^{\mu\rho}}+\sqrt{-g}\leri{g^{\mu\nu}T\indices{^\tau_{\lambda\tau}}-\delta\indices{^\nu_\lambda}T\indices{^{\tau\mu}_\tau}+T\indices{^{\nu\mu}_\lambda}}=\nonumber\\
    &=\frac{\alpha}{2}\sqrt{-g}\,\varepsilon^{\alpha\beta\gamma\nu}\leri{\mathcal{R}\indices{^\mu_{\lambda\beta\gamma}}-\frac{1}{4}\delta\indices{^\mu_\lambda}\hat{\mathcal{R}}\indices{_{\beta\gamma}}}\nabla_\alpha\theta .
    \label{equation connection general}
\end{align}
We immediately see that when the scalar field $\theta(x)$ boils down to a constant, the r.h.s. of \eqref{equation connection general} identically vanishes, and we just recover the standard equation for the connection \cite{Afonso:2017bxr,Dadhich:2012htv}. Keeping $\theta(x)$ generic, however, we can contract \eqref{equation connection general} with $\delta\indices{^\lambda_\nu},g_{\mu\nu}$ and $\varepsilon\indices{_{\rho\nu\mu}^\lambda}$, and we can extract the three equations for the four vector components of the connection $T_\mu,S_\mu,Q_\mu,P_\mu$ (see Appendix~\ref{appendix A}), i.e.
\begin{align}
    3P_\mu-\frac{3}{2}Q_\mu-2T_\mu&=\frac{\alpha}{2}\leri{\varepsilon^{\alpha\beta\gamma\delta}\mathcal{R}\indices{_{\mu\beta\gamma\delta}}+\frac{1}{4}\varepsilon\indices{_\mu^{\alpha\beta\gamma}}\hat{\mathcal{R}}\indices{_{\beta\gamma}}}\nabla_\alpha\theta,
    \label{equation component connection 1}\\
    P_\mu+\frac{1}{2}Q_\mu+2T_\mu&=\frac{\alpha}{2}\leri{\varepsilon^{\alpha\beta\gamma\delta}\mathcal{R}\indices{_{\beta\mu\gamma\delta}}+\frac{1}{4}\varepsilon\indices{_\mu^{\alpha\beta\gamma}}\hat{\mathcal{R}}\indices{_{\beta\gamma}}}\nabla_\alpha\theta,
    \label{equation component connection 3}\\
    S_\mu&=-\alpha\leri{\delta\indices{^\rho_\mu}\mathcal{R}-\mathcal{R}\indices{^\rho_\mu}-\mathcal{R}\indices{^{\rho\sigma}_{\mu\sigma}}}\nabla_\rho\theta \ .
    \label{equation component connection 4}
\end{align}
{Contraction of (\ref{equation connection general}) with $\delta\indices{^\lambda_\mu}$ results in the identity $0=0$, stemming from the invariance under projective transformations of \eqref{action CS}.} We can then extract the purely tensor part of \eqref{equation connection general}, which taking into account the definition of nonmetricity and torsion can be rewritten as
\begin{equation}
\begin{split}
q_{\nu\mu\lambda}-\Omega_{\lambda\mu\nu}&=
\frac{2}{3}\leri{T_\mu g_{\nu\lambda}-T_\nu g_{\mu\lambda}}-\dfrac{1}{6} \varepsilon_{\nu\mu\lambda\sigma}S^{\sigma}+\\
&-\frac{1}{9}\leri{g_{\mu\nu}\leri{2Q_\lambda+P_{\lambda}}-g_{\nu\lambda}\leri{4Q_\mu-7P_\mu}-g_{\mu\lambda}\leri{-\frac{1}{2}Q_\nu+2P_\nu}}\\
&+\frac{\alpha}{2}\,\varepsilon\indices{^{\alpha\beta\gamma}_\nu}\leri{\mathcal{R}\indices{_{\mu\lambda\beta\gamma}}-\frac{1}{4}g\indices{_{\mu\lambda}}\hat{\mathcal{R}}\indices{_{\beta\gamma}}}\nabla_\alpha\theta.
\label{equation connection tensor part general}
\end{split}
\end{equation}
Given that under \eqref{projective} torsion and nonmetricity vectors transform as
\begin{align}
    &T^\rho\rightarrow\tilde{T}^\rho= T^\rho-3\xi^{\rho},
    \label{transformation components projective t}\\
    &S^\rho\rightarrow\tilde{S}^\rho= S^\rho,\\
    &Q^\rho\rightarrow\tilde{Q}^\rho= Q^\rho+8\xi^{\rho},\\
    &P^\rho\rightarrow\tilde{P}^\rho= P^\rho+2\xi^{\rho},
    \label{transformation components projective}
\end{align}
we can exploit the freedom offered by the projective mode $\xi_\mu$ to get rid of one among $T_\mu,Q_\mu,P_\mu$, in order to close the system \eqref{equation component connection 1}-\eqref{equation component connection 4}, which in so doing turns out to be characterized by only three unknown vector quantities. In particular, we choose to disregard the trace part of the torsion, i.e. we set $\tilde{T}_\mu=0$ (or equivalently $\xi_\mu=T_\mu/3$), so that we are left with the system:
\begin{align}
    3P_\mu-\frac{3}{2}Q_\mu&=\frac{\alpha}{2}\leri{\varepsilon^{\alpha\beta\gamma\delta}\mathcal{R}\indices{_{\mu\beta\gamma\delta}}+\frac{1}{4}\varepsilon\indices{_\mu^{\alpha\beta\gamma}}\hat{\mathcal{R}}\indices{_{\beta\gamma}}}\nabla_\alpha\theta,
    \label{equation component connection 1 proj}\\
    P_\mu+\frac{1}{2}Q_\mu&=\frac{\alpha}{2}\leri{\varepsilon^{\alpha\beta\gamma\delta}\mathcal{R}\indices{_{\beta\mu\gamma\delta}}+\frac{1}{4}\varepsilon\indices{_\mu^{\alpha\beta\gamma}}\hat{\mathcal{R}}\indices{_{\beta\gamma}}}\nabla_\alpha\theta,
    \label{equation component connection 3 proj}\\
    S_\mu&=-\alpha\leri{\delta\indices{^\rho_\mu}\mathcal{R}-\mathcal{R}\indices{^\rho_\mu}-\mathcal{R}\indices{^{\rho\sigma}_{\mu\sigma}}}\nabla_\rho\theta,
    \label{equation component connection s proj}
\end{align}
where we dropped the tilde notation for the sake of clarity. Then, by summing and subtracting \eqref{equation component connection 1 proj}-\eqref{equation component connection 3 proj}, we can extract the symmetric and anti-symmetric part of the Riemann tensor in its first two indices (up to a contraction with the metric), i.e.
\begin{align}
    4P_\mu-Q_\mu&=\frac{\alpha}{2}\leri{\varepsilon^{\alpha\beta\gamma\delta}(\mathcal{R}\indices{_{\mu\beta\gamma\delta}}+\mathcal{R}\indices{_{\beta\mu\gamma\delta}})+\frac{1}{2}\varepsilon\indices{_\mu^{\alpha\beta\gamma}}\hat{\mathcal{R}}\indices{_{\beta\gamma}}}\nabla_\alpha\theta,
    \label{symmetric part}\\
    P_\mu-Q_\mu&=\frac{\alpha}{4}\varepsilon^{\alpha\beta\gamma\delta}(\mathcal{R}\indices{_{\mu\beta\gamma\delta}}-\mathcal{R}\indices{_{\beta\mu\gamma\delta}})\nabla_\alpha\theta.
    \label{antisymmetric part}
\end{align}
Again, we stress the fact that for constant $\theta$ we just obtain the trivial solution $Q_\mu=P_\mu=S_\mu=0$, which plugged into \eqref{equation connection tensor part general} leads to the vanishing of the purely tensor part of torsion and nonmetricity as well, as it can be appreciated by looking at \eqref{equation connection tensor part general} once we consider \eqref{symmetric part} and \eqref{antisymmetric part}, i.e.
\begin{equation}
\begin{split}
q_{\nu\mu\lambda}-\Omega_{\lambda\mu\nu}&=\frac{\alpha}{2}\,\varepsilon\indices{^{\alpha\beta\gamma}_\nu}\leri{\mathcal{R}\indices{_{\mu\lambda\beta\gamma}}-\frac{1}{4}g\indices{_{\mu\lambda}}\hat{\mathcal{R}}\indices{_{\beta\gamma}}}\nabla_\alpha\theta
-\dfrac{1}{6} \varepsilon_{\nu\mu\lambda\sigma}S^{\sigma}+\\
&-\frac{1}{9}\leri{g_{\mu\nu}\leri{2Q_\lambda+P_{\lambda}}-g_{\nu\lambda}\leri{4Q_\mu-7P_\mu}-g_{\mu\lambda}\leri{-\frac{1}{2}Q_\nu+2P_\nu}}.
\label{equation connection tensor part general solved}
\end{split}
\end{equation}

\section{Perturbative solution of the connection}\label{sec4}
In general, determining exact solutions for the system \eqref{equation component connection s proj}, \eqref{symmetric part} and \eqref{antisymmetric part} is a highly nontrivial task, so that it can be instructive to look for perturbative solutions obtained by expanding the scalar field $\theta(x)$. More precisely, we suppose for the scalar field $\theta(x)$ the perturbative expansion
\begin{equation}
    \theta(x)=\bar{\theta}+\tensor[^{(1)}]{\theta}{}(x)+\tensor[^{(2)}]{\theta}{}(x)+\dots,
\end{equation}
where $\bar{\theta}$ is the background value which we assume constant. Such an assumption is consistent with the idea that the metric-affine structure is dynamically generated by the non trivial behaviour of $\theta(x)$. Moreover, in order for the perturbative analysis to be consistent, we have to take into account the expansion of the metric field as well, which can be rewritten as $g_{\mu\nu}=\bar{g}_{\mu\nu}+h_{\mu\nu}$, where $\bar{g}_{\mu\nu}$ represents some background metric and $h_{\mu\nu}$ the tensor perturbation. In particular, by looking at \eqref{action CS}, we easily see that, being the CS correction purely affine, the metric $\bar{g}_{\mu\nu}$ is not completely arbitrary, but has to be selected among the possible GR solutions, as is clear from the lowest order of the metric field equations \eqref{equation background metric}. For a vanishing stress-energy tensor, this allows us to choose any known vacuum solution of GR. For this reason, when we focus on static and spherically symmetric spacetimes in sec.~\ref{sec6}, we will consider the Schwarzschild solution as our reference background.\\ Leaving for the time being $\bar{g}_{\mu\nu}$ as general as possible, we insert the expansions for the metric and scalar fields in \eqref{equation component connection s proj}-\eqref{antisymmetric part}, which at first order completely determine the affine structure solely in terms of the background  Riemann tensor $\bar{R}\indices{^\rho_{\mu\sigma\nu}}(\bar{g})$ and the scalar perturbation $\frstheta\equiv\delta\theta$. This is caused by the fact that the scalar field only appears with its derivatives, preventing the appearance of any bare couplings between $\bar{\theta}$ and $h_{\mu\nu}$, which are only expected to arise at higher orders. 
\\The perturbation equations then lead to 
\begin{equation}
    \tensor[^{(1)}]{P}{_\mu}=\tensor[^{(1)}]{Q}{_\mu}=0,
    \label{solution nonmetriciy linear}
\end{equation}
as it follows from \eqref{symmetric part} and 
\eqref{antisymmetric part} once we take into account the Bianchi identity and the property $\bar{R}_{\mu\nu\rho\sigma}(\bar{g})=-\bar{R}_{\nu\mu\rho\sigma}(\bar{g})$. We note that such a trivial structure for nonmetricity ultimately depends on the fact that the contribution of the homothetic curvature to the modified CS term is at least of second order in perturbation, being the homothetic curvature tensor proportional to the derivative of the nonmetricity trace, i.e. to the Weyl vector (see appendix \ref{appendix A}). It might seem, therefore, that the initial choice of $B=-\frac{1}{4}$ had a bit of arbitrariness, since it appears to be eventually irrelevant in determining \eqref{solution nonmetriciy linear}. We remark, however, that for $B\neq-\frac{1}{4}$ the contraction of \eqref{equation connection general} with $\delta\indices{^\lambda_\mu}$ does not lead to the identity $0=0$, but rather to the additional constraint:
\begin{equation}
      \epsilon^{\alpha\mu\beta\gamma}\hat{\mathcal{R}}\indices{_{\beta\gamma}}\nabla_\alpha\theta=\epsilon^{\alpha\mu\beta\gamma}\partial_\alpha\theta\partial_\beta Q_\gamma =0,
    \label{breaking projective}
\end{equation}
which allows for the Weyl vector the more general form $Q_\mu=\partial_\mu\varphi$, with $\varphi$ an undetermined scalar field. This degree of arbitrariness which the theory results to be endowed with, can be traced back to the fact that, now, the breaking of the projective invariance does not allow us to get rid of the trace $T_\mu$ of the torsion, so that we have to deal directly with \eqref{equation component connection 1}-\eqref{equation component connection 3} (the equation for the axial part of torsion is insensitive), which lead us to the set of linearized solutions:
\begin{align}
    &\tensor[^{(1)}]{P}{_\mu}=\frac{1}{4}\partial_\mu\varphi,\\
    &\tensor[^{(1)}]{T}{_\mu}=-\frac{\alpha}{4}\varepsilon^{\alpha\beta\gamma\delta}\bar{R}\indices{_{\mu\beta\gamma\delta}}\nabla_\alpha\delta\theta-\frac{1}{2}\partial_\mu\varphi.
\end{align}
The issue about the possibility that also projective breaking scenarios could be characterized by stable dynamics will be the subject of a forthcoming work \cite{supercazzola}, and here we focus on the specific case of $B=-\frac{1}{4}$.

Then, going back to \eqref{equation component connection s proj}, the solution for the axial trace of the torsion is
\begin{equation}
    \tensor[^{(1)}]{S}{_\mu}=2\alpha\, \bar{G}_{\rho\mu}(\bar{g})\nabla^\rho\delta\theta,
    \label{solution first order S}
\end{equation}
while the rank-3 components $q_{\rho\mu\nu},\,\Omega_{\rho\mu\nu}$ can be determined from \eqref{equation connection general}, which at the first order simply reduces to
\begin{equation}
    \tensor[^{(1)}]{q}{_{\nu\mu\lambda}}- \tensor[^{(1)}]{\Omega}{_{\lambda\mu\nu}}=\frac{\alpha}{2}\varepsilon\indices{^{\alpha\beta\gamma}_\nu} \bar{R}_{\mu\lambda\beta\gamma}\nabla_\alpha\delta\theta-\frac{\alpha}{3}\varepsilon_{\nu\mu\lambda\rho}\bar{G}^{\sigma\rho}(\bar{g})\nabla_\sigma\delta\theta.
    \label{equation 3-rank}
\end{equation}
Then, taking the symmetric part of \eqref{equation 3-rank} in the indices $\mu,\,\lambda$ and recalling the property $q_{\nu\mu\lambda}=-q_{\nu\lambda\mu}$, we get
\begin{equation}
    \tensor[^{(1)}]{\Omega}{_{\mu\lambda\nu}}=-\tensor[^{(1)}]{\Omega}{_{\lambda\mu\nu}}.
    \label{antisymmetry omega}
\end{equation}
Now, since $\Omega_{\mu\lambda\nu}$ is also symmetric in the last two indices, it can be verified that \eqref{antisymmetry omega} enforces the tensor $\tensor[^{(1)}]{\Omega}{_{\mu\lambda\nu}}$ to identically vanish. We are left, therefore, with
\begin{equation}
     \tensor[^{(1)}]{q}{_{\nu\mu\lambda}}=\frac{\alpha}{2}\varepsilon\indices{^{\alpha\beta\gamma}_\nu}\bar{R}_{\mu\lambda\beta\gamma}\nabla_\alpha\delta\theta-\frac{\alpha}{3}\varepsilon_{\nu\mu\lambda\rho} \bar{G}^{\sigma\rho}\nabla_\sigma\delta\theta,
\end{equation}
which a straightforward computation shows to fulfill the requirement $\varepsilon^{\nu\mu\lambda\tau}\tensor[^{(1)}]{q}{_{\nu\mu\lambda}}=0$.
Then, at the first order in $\nabla_\mu\delta\theta$ the nonmetricity is completely vanishing and solutions are only characterized by torsion, whose form is
\begin{equation}
\begin{split}
     \tensor[^{(1)}]{T}{_{\rho\mu\nu}}=\frac{\alpha}{2}\varepsilon\indices{^{\alpha\beta\gamma}_\rho} \bar{R}_{\mu\nu\beta\gamma}\nabla_\alpha\delta\theta.
     \label{solution first order torsion}
\end{split}
\end{equation}
It is important to note that this solution does not depend on the specific scalar or metric background chosen. As a result, if inserted back in the linearized equations for $h_{\mu\nu}$ and $\delta\theta$, one obtains a purely metric effective theory, where there is still freedom to select the specific background metric on which the dynamics can be studied.

\section{Metric and scalar field equations}\label{sec5}
Now, equipped with the solution for the affine structure \eqref{solution first order torsion}, we can look at the corrections due to the non Riemannian terms in the equation for the metric field, whose nonperturbative form can be easily obtained from \eqref{action CS} and reads
\begin{equation}
    \mathcal{R}_{(\mu\nu)}-\frac{1}{2}g_{\mu\nu}\mathcal{R}=\kappa T_{\mu\nu},
    \label{equation metric exact}
\end{equation}
where the stress energy tensor includes contributions from matter and the scalar field itself. Then, at the lowest order in perturbation we get the equation for the background metric $\bar{g}_{\mu\nu}$, i.e.
\begin{equation}
    \bar{G}_{\mu\nu}=\kappa \bar{T}_{\mu\nu},
    \label{equation background metric}
\end{equation}
and we immediately see that for vacuum solutions where $\bar{T}_{\mu\nu}=0$, the axial trace $S_\mu$ (eq. \eqref{solution first order S}) identically vanishes and we are only left with the tensor part of torsion.
Next, by considering the decomposition of the Riemann tensor in its metric and affine parts \eqref{perturbative expansion riemann} and the expansions $g_{\mu\nu}=\bar{g}_{\mu\nu}+h_{\mu\nu}$, $\theta=\bar{\theta}+\delta\theta$, the linearized equation for the metric takes the form
\begin{equation}
    \tensor[^{(1)}]{G}{_{\mu\nu}}+\tensor[^{(1)}]{C}{_{\mu\nu}}=\kappa \tensor[^{(1)}]{T}{_{\mu\nu}},
    \label{perturbed metric eq}
\end{equation}
where $\tensor[^{(1)}]{T}{_{\mu\nu}}$ now includes only terms depending on the matter, since the scalar field contributions are at least of second order in perturbations. Furthermore, we defined the perturbed Einstein tensor
\begin{equation}
        \tensor[^{(1)}]{G}{_{\mu\nu}}\equiv 2 \bar{\nabla}_\alpha \bar{\nabla}_{(\mu} h\indices{_{\nu)}^\alpha}-\bar{g}_{\mu\nu}\leri{\bar{\nabla}_\alpha \bar{\nabla}_\beta-\bar{R}_{\alpha\beta} } h^{\alpha\beta}- \leri{\bar{\Box}+ \bar{R}}h_{\mu\nu}+  \leri{\bar{g}_{\mu\nu}\bar{\Box} - \bar{\nabla}_\mu \bar{\nabla}_\nu} h,
\end{equation}
and we introduced the modified C-tensor (see \cite{PhysRevD.81.124021} for a comparison)
\begin{equation}
    \tensor[^{(1)}]{C}{_{\mu\nu}}\equiv  -\bar{\nabla}^\rho\tensor[^{(1)}]{T}{_{(\mu|\rho|\nu)}} = -\alpha \varepsilon_{(\mu|\alpha\gamma\delta}\leri{\bar{R}\indices{_{|\nu)\beta}^{\gamma\delta}}\bar{\nabla}^\beta  \bar{\nabla}^\alpha \delta\theta +\bar{\nabla}_\beta \bar{R}\indices{_{|\nu)}^{\beta\gamma\delta}}\bar{\nabla}^\alpha \delta\theta}.
    \label{cotton tensor first order}
\end{equation}
It is worth noting that the form of $\tensor[^{(1)}]{C}{_{\mu\nu}}$ does not coincide with its analogous in the purely metric approach, as discussed in Sec.~\ref{sec: metric CS comparison}. Then, when we vary \eqref{action CS} with respect to the scalar field $\theta(x)$, we obtain 
\begin{equation}
    \beta\Box\theta + \frac{\alpha}{8}    \varepsilon\indices{^{\mu\nu\rho\sigma}}\leri{\mathcal{R}\indices{^\alpha_{\beta\mu\nu}}\mathcal{R}\indices{^\beta_{\alpha\rho\sigma}}-\frac{1}{4}\mathcal{R}\indices{^\alpha_{\alpha\mu\nu}}\mathcal{R}\indices{^\beta_{\beta\rho\sigma}}}=0.
    \label{equation scalar field nonperturb}
\end{equation}
The linearization of the above equation on an arbitrary background results in a quite involved expression and we refer the reader to Appendix \ref{appendix B} for its explicit form. We just note, here, that the limit $\beta\to 0$ does not correspond necessarily to deprive the scalar perturbation of a proper dynamics, since by virtue of \eqref{solution first order torsion} it can be always generated by the scalar field derivatives hidden in the Riemann contractions of \eqref{equation scalar field nonperturb}. Even if we will discuss in more detail this property in Section \ref{sec6}, here we just point out that in the standard metric approach the case $\beta=0$ is non dynamical and the theory constrained by the condition
\begin{equation}
    \varepsilon\indices{^{\mu\nu\rho\sigma}}R\indices{^\alpha_{\beta\mu\nu}}R\indices{^\beta_{\alpha\rho\sigma}}=0,
    \label{pontryagin constraint metric}
\end{equation}
which can prevent some geometric configuration to be actually feasible. In our case, instead, the metric-affine generalization of \eqref{pontryagin constraint metric} results in a larger variety of possible dynamical solutions, and for $\beta=0$ the theory is still well behaved.

\section{Quasinormal modes for Schwarzschild black 
holes}\label{sec6}
\noindent 
Since by virtue of \eqref{equation connection general} a constant scalar field $\theta(x)\equiv \bar{\theta}$ implies vanishing torsion and nonmetricity, the metric and scalar field equations reduce in this case to Einstein equations with the additional condition \eqref{pontryagin constraint metric}. This implies that every GR solution satisfying \eqref{pontryagin constraint metric} will be a solution of metric-affine CSMG as well. In particular, the Schwarzschild metric with $\theta(x)\equiv \bar{\theta}$ and vanishing torsion and nonmetricity is an exact solution of the present theory.
Therefore, we can analyze the evolution in vacuum of metric and scalar perturbations, according to the procedure described in Sec.~\ref{sec4}, by selecting the Schwarzschild solution as  our background metric, i.e.
\begin{equation}
    \bar{g}_{\mu\nu}dx^\mu dx^\nu = - f(r) dt^2 + f^{-1}(r) dr^2 + r^2 d\vartheta^2 + r^2 \sin^2\vartheta d\varphi^2,
\end{equation}
with $f(r) = 1-2m/r$. Then, equation \eqref{equation scalar field nonperturb} becomes (see appendix \ref{appendix A})
\begin{equation}
    \leri{\mathcal{S}_1(\alpha,\beta;\bar{g}_{\rho\sigma})+\mathcal{S}_2(\alpha;\bar{g}_{\rho\sigma})+\mathcal{S}_3(\alpha;\bar{g}_{\rho\sigma})}\delta\theta+\mathcal{S}_4^{\mu\nu}(\alpha;\bar{g}_{\rho\sigma})h_{\mu\nu}=0,
    \label{perturbed scalar eq}
\end{equation}
where
\begin{align}
    \mathcal{S}_1(\alpha,\beta;\bar{g}_{\rho\sigma})&\equiv \leri{\beta+\alpha^2\leri{\frac{3}{4} \bar{R}^{\mu\nu\rho\sigma}\bar{R}_{\mu\nu\rho\sigma}}}\bar{\Box},\\
    \mathcal{S}_2(\alpha;\bar{g}_{\rho\sigma})&\equiv\frac{\alpha^2}{4}\leri{ 3 \bar{R}^{\mu\nu\rho\sigma}\bar{\nabla}_\alpha \bar{R}_{\mu\nu\rho\sigma} + 2\bar{R}^{\mu\nu\rho\sigma}\bar{\nabla}_\nu\bar{R}_{\alpha\mu\rho\sigma}}\bar{\nabla}^\alpha,\\
    \mathcal{S}_3(\alpha;\bar{g}_{\rho\sigma})&\equiv-2 \alpha^2 \bar{R}_{\sigma\mu\nu\rho}\bar{R}\indices{_\tau^{\mu\nu\rho}}\bar{\nabla}^\sigma\bar{\nabla}^\tau,\\
 \mathcal{S}_4^{\mu\nu}(\alpha;\bar{g}_{\rho\sigma})&\equiv\alpha\,^{*}\bar{R}\indices{^\mu_\beta^\nu_\delta}\bar{\nabla}^\delta\bar{\nabla}^\beta,
\end{align}
having used \eqref{solution first order torsion} and \eqref{perturbative expansion riemann}. Due to the spherical symmetry of the problem, the scalar field perturbation can be decomposed in standard spherical harmonics $Y^{lm}(\vartheta,\varphi)$ as
\begin{equation}
    \delta\theta = \frac{\Theta(r)}{r}Y^{lm}(\vartheta,\varphi)e^{-i\omega t}.\label{scalar decomposition}
\end{equation}
The decomposition of the metric perturbation is more involved and requires the use of tensor spherical harmonics. Since we are interested in the study of the black hole quasinormal modes, it is convenient to adopt the Regge-Wheeler gauge \cite{Maggiore:2018sht}, in which the expression of the metric perturbation simplifies to
\begin{equation}
   h_{\mu\nu} =  \begin{pmatrix}
H_0^{lm}Y^{lm} & H_1^{lm}Y^{lm} & h_0^{lm}S_\vartheta^{lm} & h_0^{lm}S_\varphi^{lm}\\
* & H_2^{lm}Y^{lm} & h_1^{lm}S_\vartheta^{lm} & h_1^{lm}S_\varphi^{lm}\\
* & * & r^2 K^{lm}Y^{lm} & 0\\
* & * & 0 & r^2 \sin^2\vartheta K^{lm}Y^{lm}
\end{pmatrix}e^{-i\omega t},
\label{metric decomposition}
\end{equation}
where asterisks denote components obtained by symmetry and we have defined
\begin{align}
    S_\vartheta^{lm} &= - \csc{\vartheta} \partial_\varphi Y^{lm},\\
    S_\varphi^{lm} &=\sin{\vartheta} \partial_\vartheta Y^{lm}.
\end{align}
Here, the functions $H_0,H_1,H_2,K$ and $h_0,h_1$ describe polar and axial perturbations, respectively, and they only depend on the radial coordinate $r$.\\
Substituting \eqref{scalar decomposition} and \eqref{metric decomposition} into \eqref{perturbed metric eq}, it can be shown that the resulting equations for the metric functions are equivalent to their analogue in metric {CSMG} \cite{PhysRevD.80.064008,PhysRevD.81.089903,PhysRevD.81.124021}. Regarding the scalar field equation instead, differences due to the affine structure are only present in terms involving the scalar field perturbation, while the last term in \eqref{perturbed scalar eq} produces the same contributions as in the metric case. This implies that the decoupling of polar and axial perturbations observed in metric {CSMG} \cite{PhysRevD.80.064008,PhysRevD.81.089903,PhysRevD.81.124021}, holds also in the affine case. In particular, the polar perturbations are not modified by the presence of the additional scalar field, hence from now on we focus on pure axial modes.

The equations describing axial perturbations are derived from the $t\varphi$, $r\varphi$ and $\vartheta\varphi$ components of \eqref{perturbed metric eq}, which read
\begin{align}
    E_1&\equiv h_0'' +i\omega \leri{\partial_r +\frac{2}{r}}h_1 + \leri{\frac{2f'}{rf} - \frac{l(l+1)}{r^2 f}}h_0 - \frac{6\alpha m }{r^4}\leri{\Theta' - \frac{2}{r}\Theta} =0,\\
    E_2&\equiv -\omega^2 h_1 +i\omega \leri{\partial_r -\frac{2}{r}}h_0 + \frac{f(l+2)(l-1)}{r^2}h_1 - \frac{6\alpha m i \omega}{r^4}\Theta=0,\\
    E_3 &\equiv \frac{i\omega }{f} h_0 + \partial_r(f h_1)=0.
\end{align}
These equations are not all independent since the following relation holds \cite{PhysRevD.80.064008}
\begin{equation}
    E_1 + \frac{f r^4}{i\omega}(E_2/r^2)' - \frac{(l+2)(l-1)r}{i\omega} E_3 =0.
\end{equation}
Moreover, one can solve $E_3=0$ for $h_0$ and substitute the result into $E_2$, yielding an equation for $h_1$, which reads
\begin{equation}
    \frac{d^2Q}{dr_*^2} + \left[ \omega^2 -f\left(\frac{l(l+1)}{r^2} - \frac{6m}{r^3}\right)\right]Q = -\frac{6 \alpha m i \omega}{r^5} f \Theta,
\end{equation}
where we defined $Q=fh_1/r$ and employed the tortoise coordinate $r_*= r + 2m\ln(r/2m - 1)$.
Then, using $E_2=0$ to express $(\partial_r-2/r)h_0$ in terms of $h_1$ and $\Theta$, the equation for the scalar perturbation eventually reads as
\begin{align}
    &\leri{\beta + \frac{12\alpha^2m^2}{r^6}}\leri{\frac{d^2\Theta}{dr_*^2} + \left[ \omega^2 -f\left(\frac{l(l+1)}{r^2} + \frac{2m}{r^3}\right)\right]\Theta}- \frac{72\alpha^2 m^2 }{r^7}f \frac{d\Theta}{dr_*}\nonumber\\
    & + \frac{36\alpha^2 m^2}{r^8}f(2f-l(l+1))\Theta = \frac{6\alpha m }{-i\omega r^5}\frac{(l+2)!}{(l-2)!} f Q.\label{scalar mode eq}
\end{align}
The last two equations form a set of coupled second order differential equations describing the evolution of (odd parity) metric and scalar perturbations on a Schwarzschild background.
In order to numerically integrate these equations and extract the QNMs frequencies, it is convenient to proceed in the following way \cite{PhysRevD.49.883,PhysRevD.63.084001,PhysRevD.70.064025}. First, the scalar field redefinition $\Theta\rightarrow\Theta/\alpha$, together with the replacement $\beta\rightarrow \alpha^2 \beta$, allow to set $\alpha=1$ in the above equations. Then, defining $\Psi=iQ/\omega$ and introducing the light-cone variables $u=t-r_*$ and $v=t+r_*$, we get
\begin{align}
    4\frac{\partial^2\Psi}{\partial u \partial v} + V_1(r) \Psi &= V_2(r) \Theta,\label{eqs light cone tensor}\\
    4 W_1(r) \frac{\partial^2\Theta}{\partial u \partial v} + W_2(r) \leri{\frac{\partial \Theta}{\partial u} - \frac{\partial \Theta}{\partial v}} + W_3(r) \Theta &= W_4(r) \Psi,
    \label{eqs light cone scalar}
\end{align}
where we defined the potentials as
\begin{align}
    V_1(r) &= f(r) \leri{\frac{l(l+1)}{r^2} - \frac{6m}{r^3}},\\
    V_2(r) &= -\frac{6m}{r^5} f(r),\\
    W_1(r) &= \beta + \frac{12m^2}{r^6},\\
    W_2(r) &= -\frac{72m^2}{r^7}f(r),\\
    W_3(r) &= f(r)\left(\beta + \frac{12m^2}{r^6}\right)\left(\frac{l(l+1)}{r^2} + \frac{2m}{r^3}\right) - \frac{36 m^2f(r)}{r^8}\leri{2f(r) - l(l+1)},\\
    W_4(r) &= -\frac{6m}{r^5}\frac{(l+2)!}{(l-2)!}f(r).
\end{align}
Note that $W_1$ is always non vanishing and positive for $\beta\geq 0$. Since the latter rules the kinetic term in the action, scalar ghost instabilities are expected to arise for $\beta<0$. For this reason, in the following we only focus on the positive branch.

To set up the numerical integration, the $u-v$ plane is discretized by a lattice spacing $\Delta$ and the equations are written in their discretized versions as
\begin{align}
    \Psi_N &= \Psi_W + \Psi_E - \Psi_S + \frac{\Delta^2}{2}\left[ V_1(r_c) \leri{\Psi_W + \Psi_E} - V_2(r_c) \leri{\Theta_W + \Theta_E} \right],\label{discrete eq metric}\\
    \Theta_N &= \Theta_W + \Theta_E - \Theta_S + \Delta \frac{W_2(r_c)}{W_1(r_c)} \leri{\Theta_E - \Theta_W}\nonumber \\
    &+ \frac{\Delta^2}{2}\left[ \frac{W_3(r_c)}{W_1(r_c)} \leri{\Theta_E + \Theta_W} - \frac{W_4(r_c)}{W_1(r_c)}\leri{\Psi_E + \Psi_W} \right].\label{discrete eq scalar}
\end{align}
Here, the fields with a subscript are evaluated at the points $S=(u,v)$, $W=(u+\Delta,v)$, $E=(u,v+\Delta)$ and $N=(u+\Delta,v+\Delta)$, while the potentials are computed at $r_c=(u+\Delta/2,v+\Delta/2)$. We let $(u,v)$ range between zero and $(u_{max},v_{max})$, thus covering a portion of the first quadrant on the $u-v$ plane. The boundary conditions are assigned on the two axes $u=0$ and $v=0$. In particular, following \cite{PhysRevD.81.124021} we choose vanishing perturbations on the $u$ axis, i.e. $\Psi(u,0)=\Theta(u,0)=0$, and Gaussian initial data on the $v$ axis, namely $\Psi(0,v)=\Theta(0,v)= \exp[-(v-v_c)^2/2\sigma]$. Knowing the fields at the three points $S$, $W$, and $E$, the integration starts computing the fields at $N$ by solving \eqref{discrete eq metric}-\eqref{discrete eq scalar}, and then proceeds to higher values of $v$, completing the full grid row by row. At each step the value of $r_c$ at which the potentials are evaluated is computed inverting the tortoise coordinate expression for $r$. This is done using the approximation $r\approx2m (1+\exp[(r_*-2m)/2m])$ for $r\approx 2m$ and numerically for larger values of $r$. Once the integration is complete, the field as a function of time is extracted at some constant $r_*$ as $\Psi(t)=\Psi(t-r_*,t+r_*)$.

\begin{figure}
  \centering
  \includegraphics[width=1\linewidth]{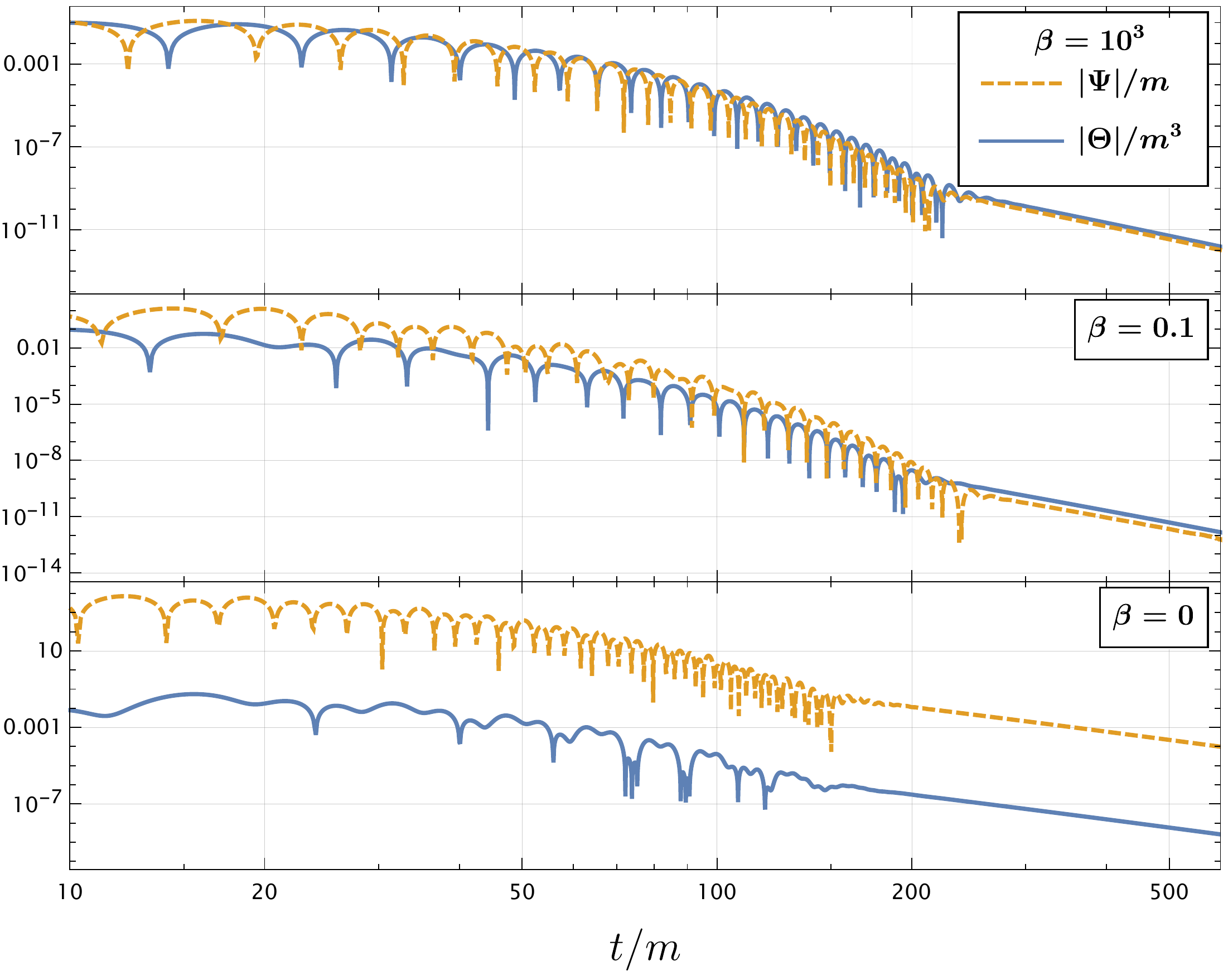}
\caption{Evolution of metric (continuous) and scalar (dashed) perturbations as a function of $t/m$ for $\beta=10^3$ (top), $\beta = 0.1$ (center) and $\beta = 0$ (bottom). Straight lines represent a power law behavior.}
\label{fig: QNMs}
\end{figure}

The results presented in this paper are obtained with the following values of the parameters: $u_{max}=v_{max}=1000$, $\Delta=0.1$, $v_c=10$, $\sigma=1$. The fields are extracted at $r_*=50m$. We performed integrations for different values of $\beta$, obtaining time series characterized by the expected behavior, namely damped oscillations ending with a late-time power law tail, as shown in Fig.~\ref{fig: QNMs}. In the next subsections we report the black hole QNMs, focusing on the lowest lying fundamental $l=2$ mode, while the behavior of the power law tails is discussed for different values of $l$.
During the integrations, for $l>2$ the perturbations reach very small amplitudes and the precision in the result is not sufficient to appreciate the power law tail. This usually happens when the amplitude becomes of order $\sim 10^{-14}$. To overcome this issue we improved the accuracy of the numerical computations using the \textit{mpmath} Python library \cite{mpmath}.
\subsection{QNMs frequencies}
\noindent The QNMs frequencies are extracted from the time series fitting the data. Depending on the value of $\beta$, either a single mode or a two modes damped oscillation is used, namely 
\begin{equation}
    y(t) = \sum_{j=1}^n A_j e^{ \Im[\omega_j] t} \cos(\Re[\omega_j] t + c_j),
\end{equation}
for $n=1$ or $n=2$, respectively. In the limit of large $\beta$ the scalar field equation decouples from the metric perturbation and reduces to the perturbed Klein-Gordon equation on a Schwarzschild background, i.e.
\begin{equation}
    \frac{d^2\Theta}{dr_*^2} + \left[ \omega^2 -f\left(\frac{l(l+1)}{r^2} + \frac{2m}{r^3}\right)\right]\Theta=0.
\end{equation}
While the equation for metric perturbations is still modified by the presence of the scalar field source on the right hand side:
\begin{equation}
    \frac{d^2\Psi}{dr_*^2} + \left[ \omega^2 -f\left(\frac{l(l+1)}{r^2} - \frac{6m}{r^3}\right)\right]\Psi = \frac{6 m  }{r^5} f \Theta.
\end{equation}
Therefore, in this regime the scalar perturbation is expected to be characterized by single mode oscillations with the same QNM frequencies as the GR case. On the other hand, the equation for the metric perturbation is always coupled to the scalar field, resulting in a superposition of the gravitational mode and the scalar one.

This behavior is confirmed by the numerical investigations performed for $\beta=100,1000$. Indeed, the metric perturbation is compatible with a two modes fit with frequencies
\begin{align}
    \omega_g &= 0.37 - i\, 0.089,\\
    \omega_s &= 0.48 - i\, 0.097,
\end{align}
while only the latter characterizes the behavior of the scalar field. Note that the above values coincide with the lowest lying ($l=2$) modes for tensor and scalar perturbations in GR, respectively.

\begin{figure}
  \centering
  \includegraphics[width=1\linewidth]{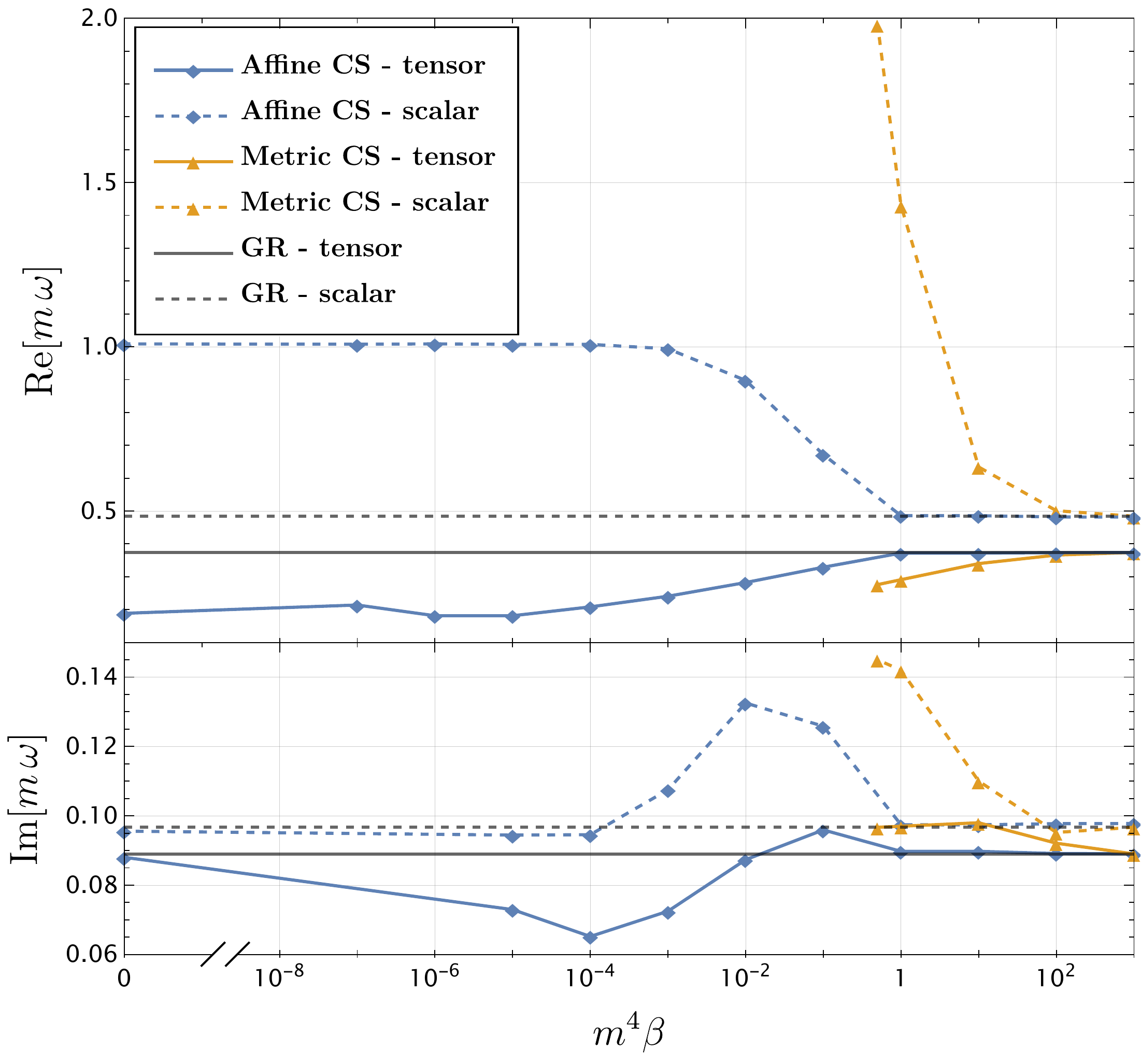}
\caption{Real (top) and imaginary (bottom) part of quasinormal frequencies of the fundamental $l=2$ tensor (continuous) and scalar (dashed) modes as a function of $\beta$, for metric (data taken from \cite{PhysRevD.81.124021}) and metric-affine {CSMG}.}
\label{fig: freq}
\end{figure}

For $10^{-1}\lesssim\beta \lesssim 10$, the numerical values of the two frequencies are almost unchanged but both perturbations start oscillating with a superposition of the two modes. This two modes behavior is also present for lower values of $\beta$, down to $\beta=0$, but for $\beta \lesssim 10^{-1}$ the frequencies deviate from the GR values as shown in Fig. \ref{fig: freq}.

When the frequencies are extracted from two modes curves, one actually obtains two values for each frequency, one from the scalar and another from the metric perturbation. Concerning the real parts, the two values obtained are always consistent with each other, while the extraction of the imaginary part is not always possible without ambiguities and we are not able to compute it for every value of $\beta$.

\subsection{Power law tails}
\begin{figure}
  \centering
  \includegraphics[width=.5\linewidth]{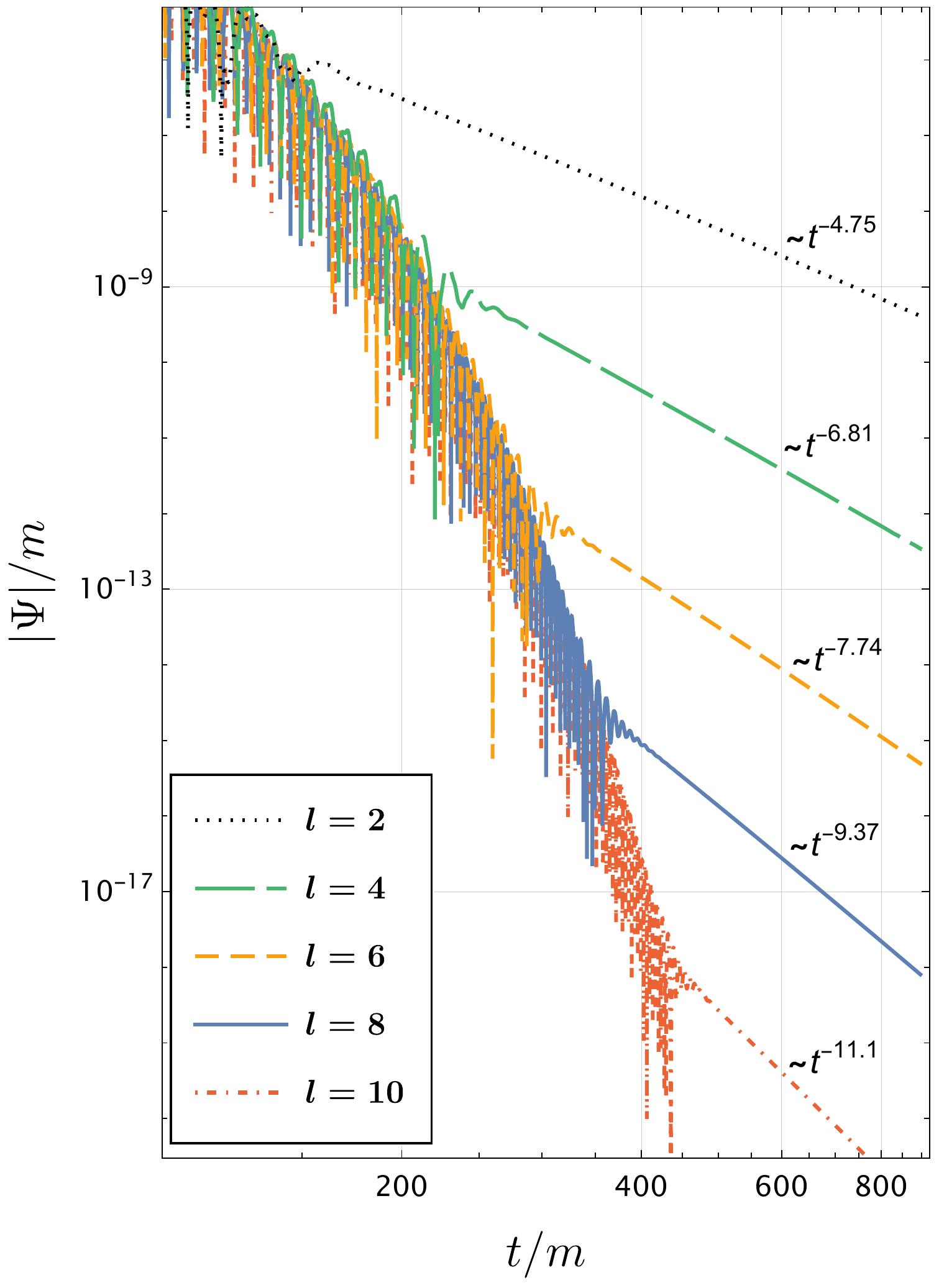}
\caption{Time evolution of the metric perturbation in the $\beta=0$ case, for different values of $l$. The behavior of scalar perturbations is qualitatively the same.}
\label{fig: tails}
\end{figure}

\noindent In each case considered the late-time evolution of the  perturbations is characterized by power law tails $\sim  t^{-\mu}$, with scalar and tensor perturbations always sharing the same value of the exponent $\mu$. In the $\beta\neq 0$ case the late-time tails are indistinguishable from GR. In particular, the exponent only depends on the angular number $l$ and is compatible with the relation $\mu=2l+3$, for each value of $\beta$. For $\beta=0$ instead, a departure from GR is observed (see Fig.~\ref{fig: tails}).

\begin{figure}
  \centering
  \includegraphics[width=1\linewidth]{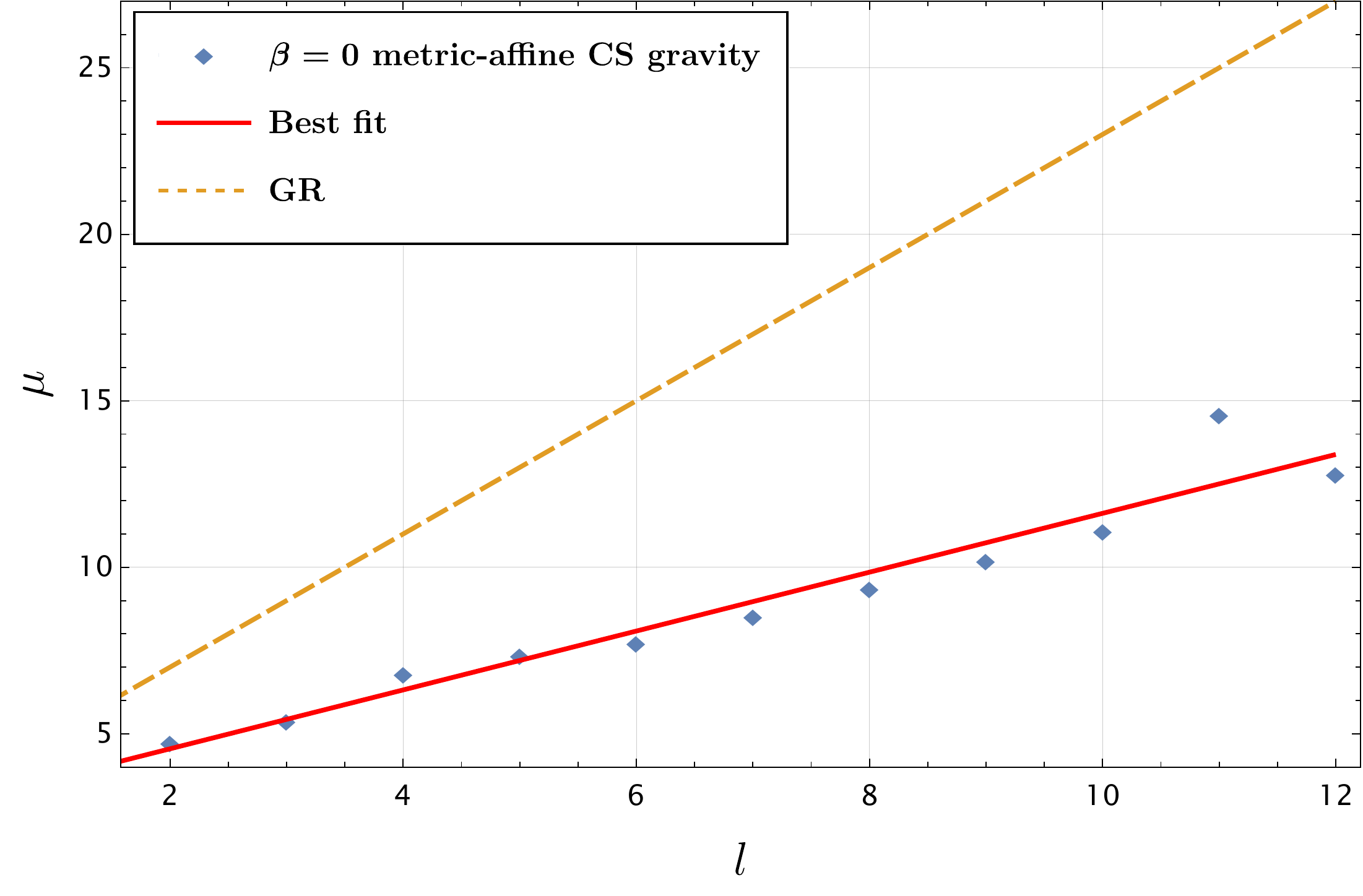}
\caption{Exponents characterizing the power law tails in the $\beta=0$ case as a function of $l$. The best fit \eqref{tails relation} (continuous) and GR case (dashed) are also shown.}
\label{fig: exponents}
\end{figure}

Power law tails obtained for $2\leq l\leq 12$ are characterized by the exponents in Fig.~\ref{fig: exponents} and Table~\ref{tab: exponents}. The relation between $\mu$ and $l$ is still linear but consistent with
\begin{equation}
    \mu = 0.884 \, l + 2.78.
    \label{tails relation}
\end{equation}
\begin{table}[b]
\centering
    \begin{tabular}{|c|c|c|c|c|c|c|c|c|c|c|c|}
    \hline
        $l$ & 2 & 3 & 4 & 5 & 6 & 7 & 8 & 9 & 10 & 11 & 12 \\
        \hline
        $\mu$ & 4.75 & 5.40 & 6.81 & 7.36 & 7.74 & 8.53 & 9.37 & 10.2 & 11.1 & 14.6 & 12.8 \\
    \hline
    \end{tabular}
    \caption{Exponents characterizing power law tails $\sim t^{-\mu}$ for different values of $l$ in the $\beta=0$ case.}
    \label{tab: exponents}
\end{table}

\subsection{Comparison with metric CSMG}\label{sec: metric CS comparison}
\noindent A comparison with dynamical {CSMG} in metric formalism can be traced considering the effective theory characterized by equations \eqref{perturbed metric eq} and \eqref{perturbed scalar eq}, where the affine structure has already been re-expressed in terms of the scalar field. A first difference arises in the expression of the C-tensor featuring the metric equation \eqref{perturbed metric eq}, which does not coincide with its analog in the purely metric approach, by virtue of the derivative acting on the Riemann tensor, which in the metric case is replaced by the Ricci tensor (Cf. with \cite{PhysRevD.81.124021}). Moreover, despite the scalar field equation \eqref{equation scalar field nonperturb} being formally equivalent at linearized level to its metric counterpart, here we actually deal with the complete Riemann tensor, including its non Riemannian contributions, which in turn depends on derivatives of the scalar field. Indeed, equation \eqref{equation scalar field nonperturb} can be expressed as \eqref{full perturbed scalar eq}, where additional $\delta\theta$ derivative terms appear explicitly.
However, even if these differences hold in the general case, they can disappear when we deal with specific background configurations, as it occurs for instance when we look at Schwarzschild spacetimes. In that case the metric equation reduces to its purely metric analogue while the scalar field equation retains the additional terms in \eqref{perturbed scalar eq}, which ultimately yield the modified structure of \eqref{scalar mode eq}.
The latter is responsible for the differences in the results obtained above with respect to the metric case. A further distinction consists on the absence of the stable exponentially decaying mode characterized by non oscillatory frequencies with $\Re[\omega]=0$ and $\Im[\omega]<0$, observed in the metric case \cite{PhysRevD.81.124021} for $\beta\lesssim 0.5$. Here we checked that only oscillatory modes with $\text{Re}[\omega]\neq 0$ are present, for $\beta$ down to $10^{-7}$.

Finally, while the late-time tails have the same behavior as in metric {CSMG} and GR for $\beta\neq 0$, the results differ in the $\beta=0$ case, which is a viable theory endowed with an additional stable scalar degree of freedom with a proper dynamical character. Contrary to metric {CSMG}, in which an observation of the late-time signal would not be able to discern the theory from GR, the $\beta=0$ version of metric-affine {CSMG} would show a distinctive signature in the exponents of the power law tails, since they would satisfy the modified relation \eqref{tails relation}.

Let us close this section with a further comparison between the metric and metric-affine versions of the theory. Metric CSMG is known to be fully consistent only in an effective field theory approach, namely in the limit of small coupling $|\alpha^2/\beta| \ll 1$. This is due to the derivative of the Ricci tensor appearing in the metric version of the C-tensor, which ultimately causes the presence of third order derivatives of the metric tensor in the field equations. Actually, in a perturbative framework at linear order they play no role because they are multiplied by derivatives of $\theta$, thus forbidding terms linear in the metric perturbation and its derivatives (if the scalar field is expanded around a constant value). However, third order derivatives may cause issues in more general settings, e.g. when discussing the initial value formulation of dynamical CSMG, where one is forced to resort to the effective field theory framework \cite{PhysRevD.91.024027}.

The situation in metric-affine CSMG is more puzzling. In this case, indeed, we can identify the C-tensor with the disformal tensor contributions stemming from \eqref{equation metric exact}, once the decomposition of the Riemann tensor in terms of $N\indices{^\rho_{\mu\nu}}$ and its metric covariant derivative is considered (see \eqref{perturbative expansion riemann}). That allows us to formulate a quite general criterion for establishing whether such terms can introduce dynamical instabilities or not: exact (non perturbative) solutions of the non-Riemannian part of the connection must contain at most first derivatives of the metric and scalar fields, in order for terms proportional to $\tensor[^{(L)}]{\nabla}{_{\sigma}}N\indices{_{\rho\mu\nu}}$ to be harmless. Clearly, the scenario analyzed in this work falls into these conditions, and the reason is rooted in the perturbative expansion at linear order, which prevents the appearance of the metric perturbation in the expression for linearized torsion, in close analogy to the metric case, as explained above.
In the general case, however, determining the exact form of torsion and nonmetricity is very challenging and we cannot exclude a priori the presence of higher order derivatives in metric-affine CSMG, even if it is not as manifest as in the metric formulation. Despite this, it is worth mentioning that a careful analysis of the equations for the connection may help in establishing which non-Riemannian components are responsible for the instabilities, so that in principle they can be removed by introducing specific symmetries in the initial formulation of the problem. In this sense, preliminary results about the isotropic and homogeneous cosmological case (Friedmann-Robertson-Walker spacetimes), where exact and nonperturbative solutions for torsion and nonmetricity can be actually derived\footnote{This will be discussed in more detail in a forthcoming work.}, seem to validate this hypothesis, offering a precious hint about a formulation of CSMG free of dynamical pathologies.

\section{Conclusions}\label{sec7}
In this paper we studied perturbations of Schwarzschild black holes in metric-affine {CSMG}, where the metric field and the connection are assumed to be independent geometric entities. In particular, we proposed to enlarge the metric formulation of the Pontryagin density with additional terms quadratic in the homothetic curvature. We demonstrated how this modification is the unique choice allowing to recover the projective invariance of the theory without spoiling the topological character of the Pontryagin density, which still holds when the pseudo-scalar field coupling with the additional terms is set to a constant value, as it happens in the standard  metric {CSMG}. Moreover, we included a standard kinetic term for the pseudo-scalar field ruled by a free parameter $\beta$.

The equations of motion for the independent connection turned out to be nonlinear and differential, so that looking for an exact solution in terms of the metric and pseudo-scalar field is prohibitive. We proceeded, therefore, by expanding the scalar field in a perturbative series and assuming the affine structure to be sourced by the latter. This allowed us to solve for the connection neglecting quadratic terms in the scalar field. Furthermore, by taking advantage of the projective symmetry, we set to zero the trace of torsion, and we ended up with an affine structure entirely determined by the axial and purely tensor part of torsion, both proportional to the pseudo-scalar field gradient. The nonmetricity tensor, instead, turned out to be vanishing at first order in the perturbation expansion. In turn, this implies that being the homothetic curvature only determined by nonmetricity, the projective invariant modification included in the action does not affect the equations at the linearized level, although it highly modifies the complete nonperturbative theory.

Then, substituting the solution for the connection back into the theory, we obtained the remaining metric and scalar field equations. The presence of a modified C-tensor in the former and of additional derivative terms in the latter, cause a discrepancy with respect to the correspondent equations in metric {CSMG}. The modified differential structure of the scalar equation guarantees a dynamical character to the pseudo-scalar perturbation even in absence of the standard kinetic term in the action ($\beta = 0$), contrary to what happens in non-dynamical metric {CSMG}. On the other hand, the discrepancy in the C-tensors disappears when specializing to a Schwarzschild background, while the pseudo-scalar field equation retains its peculiar modified structure. Employing a harmonic decomposition and working in the Regge-Wheeler gauge we solved the coupled differential equations via numerical methods, obtaining the evolution of metric and pseudo-scalar field perturbations. Because of the parity violating character of the modifications included in the action we focus only on axial, odd parity modes, which are the only ones affected.

The computation of the lowest lying ($l=2$) QNMs reveals an apparent deviation from both GR and metric {CSMG}, and the results depend on the value of the parameter $\beta$. The QNMs are always characterized by two frequencies, identifying tensor and scalar modes. In the large $\beta$ limit ($\beta \gtrsim 100$), the corresponding frequencies coincide with their GR counterpart. However, no matter how large $\beta$ is, the pseudo-scalar field acts as a source term in the metric equation and the metric perturbation always carries a signature of the pseudo-scalar degree of freedom: it evolves as a superposition of the tensor and scalar modes. On the other hand, the pseudo-scalar perturbation decouples and it exhibits single mode oscillations, identified by the scalar frequency. As $\beta$ decreases below $\beta \sim 10$, however, the two modes behavior becomes dominant in the evolution of the scalar perturbation as well.

Moreover, as shown in Fig.~\ref{fig: freq}, the frequencies deviate from GR and metric {CSMG}, smoothly reaching the values attained in the $\beta = 0$ case. The deviation from GR starts at smaller values of $\beta$, with respect to metric {CSMG}, meaning that there exists a larger range of $\beta$ values mimicking the GR results. However, the superposition of two modes is always present. As discussed in \cite{PhysRevD.81.124021}, if a signal able to resolve the two frequencies is detected, these would be interpreted as the $l=2$ tensor and scalar modes in {CSMG} or as the $l=2$ and $l=3$ tensor modes in GR (in which there are no additional scalar fields), offering a way to test the predictions of the theory. The critical quantity is the signal to noise ratio (SNR) of the gravitational waves detector \cite{PhysRevD.73.064030,PhysRevD.76.104044,Berti_2009}. In this respect, the SNR necessary to resolve the two frequencies in the $\beta=0$ affine theory is lower than the one required to disentangle the frequencies in GR or in the large $\beta$ limit of the purely metric case, due to the greater gap between the two frequencies (See Fig.~\ref{fig: freq}).

Moreover, a new feature entirely due to the affine structure of the theory with $\beta = 0$ is the modified behavior of the late-time power law tails. The exponent characterizing the tails has a linear dependence on $l$ differing from the relation found in GR and metric {CSMG}, which is recovered in our case only when $\beta \neq 0$. The discontinuous transition to the $\beta = 0$ results is due to the different asymptotic behavior of the effective potentials featuring \eqref{eqs light cone tensor}-\eqref{eqs light cone scalar}, as already observed in other scenarios \cite{PhysRevLett.74.2414,PhysRevD.52.2118}. Finally, let us remark here that integrations performed with $\beta < 0$ yield unstable, diverging modes with $\Im[\omega]>0$. This is expected since $\beta$ rules the kinetic term in the action and a change in its sign can be related to the arising of ghost instabilities.

Given the amount of information that can be extracted studying black hole perturbations, it could be an interesting perspective to extend the present analysis to non static spacetimes and non singular compact objects, possibly developing analytical methods adapted from other contexts to sustain the numerical techniques employed here.

\acknowledgments
S.B. thanks the Dept. of Theoretical Physics \& IFIC of the University of Valencia \& CSIC for their hospitality during the elaboration of this work. 
The work of F. B. has been supported by the Fondazione Angelo della Riccia grant. PJP would like to thank the Brazilian agency CAPES for financial
support (PNPD/CAPES grant, process 88887.464556/2019-00) and Department de Física Teòrica and IFIC, Universitat de València, for hospitality. This work is supported by the Spanish Grant FIS2017-84440-C2- 1-P funded by MCIN/AEI/10.13039/5011\\
00011033 “ERDF A way of making Europe”, Grant PID2020-116567GB-C21 funded by MCIN/AEI/10.13039/501100011033, the project PROMETEO/2020/079 (Generalitat Valenciana), and by  the European Union's Horizon 2020 research and innovation programme under the H2020-MSCA-RISE-2017 Grant No. FunFiCO-777740.

\appendix
\section{Metric-affine formalism}\label{appendix A}
\noindent In this appendix we review some basic notions about the metric-affine formalism we adopted throughout the paper. The Riemann tensor is defined in terms of the independent connection as:
\begin{equation}
    \mathcal{R}\indices{^\rho_{\mu\sigma\nu}}=\partial_\sigma\Gamma\indices{^\rho_{\mu\nu}}-\partial_\nu\Gamma\indices{^\rho_{\mu\sigma}}+\Gamma\indices{^\rho_{\tau\sigma}}\Gamma\indices{^\tau_{\mu\nu}}-\Gamma\indices{^\rho_{\tau\nu}}\Gamma\indices{^\tau_{\mu\sigma}},
\end{equation}
and covariant derivatives act as
\begin{equation}
    \nabla_\mu T\indices{^\rho_\sigma}=\partial_\mu T\indices{^\rho_\sigma}+\Gamma\indices{^\rho_{\lambda\mu}}T\indices{^\lambda_\sigma}-\Gamma\indices{^\lambda_{\sigma\mu}}T\indices{^\rho_\lambda}.
\end{equation}
We are considering the affine connection as general as possible, so that we can introduce torsion and nonmetricity tensors, which read respectively:
\begin{equation}
    \begin{split}
        &T\indices{^\rho_{\mu\nu}}\equiv\Gamma\indices{^\rho_{\mu\nu}}-\Gamma\indices{^\rho_{\nu\mu}},\\
        &Q\indices{_{\rho\mu\nu}}\equiv-\nabla_\rho g_{\mu\nu}.
    \end{split}
\end{equation}
In evaluating the equation of motion for the connection from \eqref{equation connection general}, we used the generalized Palatini identity
\begin{equation}
    \delta\mathcal{R}\indices{^\rho_{\mu\sigma\nu}}=\nabla_\sigma\delta\Gamma\indices{^\rho_{\mu\nu}}-\nabla_\nu\delta\Gamma\indices{^\rho_{\mu\sigma}}-T\indices{^\lambda_{\sigma\nu}}\delta\Gamma\indices{^\rho_{\mu\lambda}},
\end{equation}
and the property for vector densities
\begin{equation}
    \int d^4 x\; \nabla_\mu\leri{\sqrt{-g} V^\mu}=\int d^4x\; \partial_\mu\leri{\sqrt{-g}V^\mu}+\int d^4x\; \sqrt{-g}\;T\indices{^\rho_{\mu\rho}} V^\mu=\int d^4x\; \sqrt{-g}\;T\indices{^\rho_{\mu\rho}} V^\mu.
\end{equation}
We also rewrote torsion and nonmetricity in their irreducible parts:
\begin{align}
    &T_{\mu\nu\rho} = \dfrac{1}{3}\left(T_{\nu}g_{\mu\rho}-T_{\rho}g_{\mu\nu}\right) +\dfrac{1}{6} \varepsilon_{\mu\nu\rho\sigma}S^{\sigma} + q_{\mu\nu\rho},\label{torsion decomposition}\\
    &Q_{\rho\mu\nu}=\frac{5Q_\rho-2P_\rho}{18}g_{\mu\nu}-\frac{Q_{(\mu}g_{\nu)\rho}-4P_{(\mu}g_{\nu)\rho}}{9}+\Omega_{\rho\mu\nu}.
    \label{non metricity decomposition}
\end{align}
In particular, we introduced the trace vector
\begin{equation}
T_{\mu} \equiv T \indices{^{\nu}_{\mu\nu}},
\end{equation}
the pseudotrace axial vector
\begin{equation}
S_{\mu} \equiv \varepsilon_{\mu\nu\rho\sigma}T^{\nu\rho\sigma},
\end{equation}
and the antisymmetric tensor $q_{\mu\nu\rho}=-q_{\mu\rho\nu}$ satisfying
\begin{equation}
\varepsilon^{\mu\nu\rho\sigma} q_{\nu\rho\sigma} = 0, \qquad q\indices{^{\mu}_{\nu\mu}} = 0.
\end{equation}
While for what concerns the nonmetricity, we defined the Weyl vector
\begin{equation}
    Q_\rho=Q\indices{_\rho^\mu_\mu},
\end{equation}
the second trace
\begin{equation}
    P_\rho=Q\indices{^\mu_{\mu\rho}}=Q\indices{^\mu_{\rho\mu}},
\end{equation}
and the traceless part $\Omega_{\rho\mu\nu}$, obeying
\begin{equation}
    \Omega_{\rho\mu\nu}=\Omega_{\rho\nu\mu}.
\end{equation}
It is always possible, moreover, to rewrite the affine connection as
\begin{equation}
    \Gamma\indices{^\rho_{\mu\nu}}=L\indices{^\rho_{\mu\nu}}+N\indices{^\rho_{\mu\nu}}=L\indices{^\rho_{\mu\nu}}+K\indices{^\rho_{\mu\nu}}+D\indices{^\rho_{\mu\nu}},
    \label{christoffel contorsion disformal}
\end{equation}
where $L\indices{^\rho_{\mu\nu}}$ denotes the Christoffel symbols and the contorsion and disformal tensors are given by, respectively
\begin{align}
    &K\indices{^\rho_{\mu\nu}}=\frac{1}{2}\leri{T\indices{^\rho_{\mu\nu}}-T\indices{_\mu^\rho_\nu}-T\indices{_\nu^\rho_\mu}}=-K\indices{_\mu^\rho_{\nu}},
    \label{decomposition contorsion}\\
    &D\indices{^\rho_{\mu\nu}}=\frac{1}{2}\leri{Q\indices{_{\mu\nu}^\rho}+Q\indices{_{\nu\mu}^\rho}-Q\indices{^\rho_{\mu\nu}}}=D\indices{^\rho_{\nu\mu}}.
    \label{decomposition disformal}
\end{align}
For a generic metric-affine structure the Riemann tensor is skew-symmetric only in its last two indices, so that we can in principle take the different traces
\begin{align}
    \mathcal{R}_{\mu\nu}&\equiv\mathcal{R}\indices{^\alpha_{\mu\alpha\nu}},\\
    \hat{\mathcal{R}}_{\mu\nu}&\equiv \mathcal{R}\indices{^\alpha_{\alpha\mu\nu}}=\partial_{[\mu}Q_{\nu]},\\
    \mathcal{R}^\dag_{\mu\nu}&\equiv g_{\mu\tau}g^{\rho\sigma}\mathcal{R}\indices{^\tau_{\rho\sigma\nu}},
\end{align}
where the first one is the usual Ricci tensor, and the second one is also called homothetic curvature. In terms of the distorsion tensor the Riemann curvature can be rewritten as
\begin{equation}
        \mathcal{R}_{\mu\rho\nu\sigma}=R_{\mu\rho\nu\sigma}+\tensor[^{(L)}]{\nabla}{_{\nu}}N_{\mu\rho\sigma}-\tensor[^{(L)}]{\nabla}{_{\sigma}} N_{\mu\rho\nu}+N_{\mu\lambda\nu}N\indices{^\lambda_{\rho\sigma}}-N_{\mu\lambda\sigma}N\indices{^\lambda_{\rho\nu}},
    \label{perturbative expansion riemann}
\end{equation}
where $R\indices{^\mu_{\nu\rho\sigma}}$ and $\tensor[^{(L)}]{\nabla}{_{\mu}}$ are built from the Levi Civita connection.
\section{Scalar field equation}\label{appendix B}

Here we report the linearized equation for the field $\theta(x)$ on an arbitrary background. For the sake of clarity, it is convenient to rearrange it in the following way
\begin{equation}
    \leri{\mathcal{D}_1(\alpha,\beta;\bar{g}_{\rho\sigma})+\mathcal{D}_2(\alpha;\bar{g}_{\rho\sigma})+\mathcal{D}_3(\alpha;\bar{g}_{\rho\sigma})}\delta\theta+\mathcal{D}_4^{\mu\nu}(\alpha;\bar{g}_{\rho\sigma})h_{\mu\nu}+C(\alpha;\bar{g}_{\rho\sigma},h_{\rho\sigma})=0,
    \label{full perturbed scalar eq}
\end{equation}
where $\mathcal{D}_i$ denote differential operators acting on the field perturbations $\delta\theta$ and $h_{\mu\nu}$, and $C(\alpha;\bar{g}_{\rho\sigma},h_{\rho\sigma})$ is a function both of the background metric and the gravitational perturbation. They are given by, respectively:
\begin{align}
    \mathcal{D}_1(\alpha,\beta;\bar{g}_{\rho\sigma})&\equiv \leri{\beta+\alpha^2\leri{\frac{3}{4} \bar{R}^{\mu\nu\rho\sigma}\bar{R}_{\mu\nu\rho\sigma}-\bar{R}_{\alpha\beta}\bar{R}^{\alpha\beta}}}\bar{\Box} \equiv \leri{\beta+\alpha^2 K}\bar{\Box},\\
    \mathcal{D}_2(\alpha;\bar{g}_{\rho\sigma})&\equiv \alpha^2 \leri{\bar{R}\indices{^\mu_\sigma}\bar{R}_{\mu\tau}+2\bar{R}^{\mu\nu}\bar{R}_{\sigma\mu\tau\nu}-2\bar{R}_{\sigma\mu\nu\rho}\bar{R}\indices{_\tau^{\mu\nu\rho}}}\bar{\nabla}^\sigma\bar{\nabla}^\tau,\\
    \mathcal{D}_3(\alpha;\bar{g}_{\rho\sigma})&\equiv\frac{\alpha^2}{2}\left(\bar{\nabla}_\sigma K +\bar{R}^{\mu\nu\rho\tau}\bar{\nabla}_\nu\bar{R}_{\sigma\mu\rho\tau}-3\bar{R}\indices{_\sigma^{\tau\mu\rho}}\bar{\nabla}_\nu\bar{R}\indices{^\nu_{\tau\mu\rho}}+2\bar{R}\indices{_\sigma^\beta}\bar{\nabla}_\mu\bar{R}\indices{_\beta^\mu}\right.\nonumber\\
    &\left. +2\bar{\nabla}_\tau\leri{\bar{R}^{\mu\nu}\bar{R}\indices{_{\mu\sigma\nu}^\tau}}\right)
 \bar{\nabla}^\sigma,\\
 \mathcal{D}_4^{\mu\nu}(\alpha;\bar{g}_{\rho\sigma})&\equiv\alpha\,^{*}\bar{R}\indices{^\mu_\beta^\nu_\delta}\bar{\nabla}^\delta\bar{\nabla}^\beta,\\
 C(\alpha;\bar{g}_{\rho\sigma},h_{\rho\sigma})&\equiv\frac{\alpha}{2}\,^{*}\bar{R}\indices{_{\mu\nu\beta\delta}}\leri{h^{\alpha\beta}\bar{R}\indices{_\alpha^{\delta\mu\nu}}+ h^{\alpha\mu}\bar{R}\indices{_\alpha^{\nu\beta\delta}}-\frac{1}{4} h\indices{^\alpha_\alpha} \bar{R}^{\beta\delta\mu\nu} }.
\end{align}

\bibliographystyle{JHEP}
\bibliography{references}

\end{document}